\documentclass[%
 reprint,
twocolumn,aps,prd,showpacs,superscriptaddress,amsmath,amssymb,nofootinbib,nobibnotes]{revtex4-2}

\usepackage{graphicx}
\usepackage{dcolumn}
\usepackage{bm}
\usepackage{subfigure}
\usepackage[dvipsnames]{xcolor}
\usepackage{soul}

\begin{document}

\title{Einstein-Maxwell-dilaton neutral black holes in strong magnetic fields: \\ topological charge, shadows and lensing}

\author{Haroldo C. D. Lima Junior}
\email{haroldolima@ufpa.br} 
\affiliation{Departamento de Matem\'atica da Universidade de Aveiro and Centre for Research and Development  in Mathematics and Applications (CIDMA), Campus de Santiago, 3810-183 Aveiro, Portugal }
\affiliation{Programa de P\'os-Gradua\c{c}\~{a}o em F\'{\i}sica,
Universidade Federal do Par\'a, 66075-110, Bel\'em, PA, Brasil }

\author{Jian-Zhi Yang}
\email{jianzhi@ua.pt} 
\affiliation{Departamento de Matem\'atica da Universidade de Aveiro and Centre for Research and Development  in Mathematics and Applications (CIDMA), Campus de Santiago, 3810-183 Aveiro, Portugal }

\author{Lu\'{\i}s C. B. Crispino}
\email{crispino@ufpa.br} 
\affiliation{Programa de P\'os-Gradua\c{c}\~{a}o em F\'{\i}sica,
Universidade Federal do Par\'a, 66075-110, Bel\'em, PA, Brasil }

\author{Pedro V. P. Cunha}%
 \email{pvcunha@ua.pt}
\affiliation{Departamento de Matem\'atica da Universidade de Aveiro and Centre for Research and Development  in Mathematics and Applications (CIDMA), Campus de Santiago, 3810-183 Aveiro, Portugal }

\author{Carlos A. R. Herdeiro}
\email{herdeiro@ua.pt} 
\affiliation{Departamento de Matem\'atica da Universidade de Aveiro and Centre for Research and Development  in Mathematics and Applications (CIDMA), Campus de Santiago, 3810-183 Aveiro, Portugal }

\date{December 2021}

\begin{abstract}
The light rings (LRs) topological charge (TC) of a spacetime measures the number of stable LRs minus the number of unstable LRs. It is invariant under smooth spacetime deformations obeying fixed boundary conditions. Asymptotically flat equilibrium black holes (BHs) have, generically, TC=$-1$. In Einstein-Maxwell theory, however, the Schwarzschild-Melvin BH - describing a neutral BH immersed in a strong magnetic field - has TC$=0$. This allows the existence of \textit{BHs without LRs} and produces remarkable phenomenological features, like panoramic shadows. Here we investigate the generalised Schwarzschild-Melvin solution in Einstein-Maxwell-\textit{dilaton} theory, scanning the effect of the dilaton coupling $a$. We find that the TC changes discontinuously from TC$=0$ to TC$=-1$ precisely at the Kaluza-Klein value $a=\sqrt{3}$, when the (empty) Melvin solution corresponds to a twisted Kaluza-Klein reduction of five-dimensional flat spacetime, $i.e.$ the dilaton coupling $a$ induces a \textit{topological transition} in the TC. We relate this qualitative change to the Melvin asymptotics for different $a$. We also study the shadows and lensing of the generalised Schwarzschild-Melvin solution for different values of $a$, relating them to the TC.
\end{abstract}

\maketitle

\tableofcontents

\section{Introduction}
The light rings (LRs) topological charge (TC)~\cite{Cunha:2017qtt} has proved to be a powerful concept to establish generic properties on the existence and type of LRs around compact objects. It was originally proposed to demonstrate that any equilibrium ultracompact object, stationary and axi-symmetric, that can result from (an incomplete) gravitational collapse of a near Minkowskian spacetime must have TC$=0$~\cite{Cunha:2017qtt}. This means that LRs must come in pairs for such ultracompact objects, with one being stable and another unstable in each pair  - see also~\cite{Cardoso:2014sna} for an earlier work restricted to spherical symmetry and~\cite{Hod:2017zpi} for a discussion on the degenerate case. Thus, if a horizonless ultracompact object contains an unstable LR, as to mimic the properties of black holes (BHs), another (stable) LR must also be present. It has been argued that the latter triggers a spacetime instability~\cite{Keir:2014oka,Benomio:2018ivy}. Thus, this result has the potential to rule out ultracompact objects smoothly formed from Minkowski spacetime as dynamically viable BH mimickers. Yet, the confirmation of the suggested spacetime instability and the corresponding timescales are open questions (see~\cite{Guo:2021bcw} for a discussion in a specific example).

Another application of the TC was given in~\cite{PRLCunha:2020}, to show that, generically, asymptotically flat equilibrium BHs, even in modified gravity, have TC$=-1$. In other words, BHs have one unstable LR in excess to any other possible stable-unstable LR pairs. This is of course the case for the paradigmatic BHs of General Relativity (GR), described by the Kerr metric~\cite{Kerr:1963ud}. Several generalizations of this result, for which TC$=-1$ still holds in BH spacetimes,  have been discussed - see  $e.g.$~\cite{Wei:2020rbh,Guo:2020qwk,Koga:2020akc,Ghosh:2021txu,Lim:2021lju, PaganiniThesis}.

A qualitatively different example for the TC of a BH spacetime, however, was given in~\cite{MSBH21}. Therein a Schwarzschild BH was immersed in a strong magnetic field, described by the Melvin solution of Einstein-Maxwell theory~\cite{Melvin:1963qx}. The resulting Schwarzschild-Melvin spacetime, first discussed by Ernst~\cite{Ernst:1976mzr}, was shown to have TC$=0$. This means that, regardless of how small, but non-zero, the magnetic field $B$ is, it changes the spacetime concerning its LRs composition. At the basis of this result stands the fact that any $B\neq 0$ changes the spacetime asymptotics; thus no Melvin solution, regardless of how small $B$ may be, is a perturbation of flat spacetime. A similar status holds for a cosmological constant: no matter how small, it changes the global structure of the spacetime. It was shown in~\cite{MSBH21} that the Melvin asymptotics introduce a stable LR, which, together with the usual (unstable) Schwarzschild LR, explains TC$=0$. However, for a fixed BH mass $M$, there is critical value $B_c$, for which the two LRs merge, and for $B>B_c$ the BH spacetime has no LRs. 

It is well known that there is a close relation between the LRs, or more generically the fundamental photon orbits (FPOs)~\cite{Cunha:2017eoe}, and the edge of what is generally known as the BH \textit{shadow}~\cite{Falcke:1999pj} - see~\cite{Bardeen:1973tla,Luminet:1979nyg}. Thus, the absence of LRs for sufficiently large $BM$ in the Schwarzschild-Melvin spacetime raises the interesting question on how does this impact on the BH shadow. In fact, it was discussed in~\cite{MSBH21} that this leads to panoramic shadows, seen (almost) all around the observers' sky, along the equatorial plane.  

The Schwarzschild-Melvin example shows how the asymptotic structure may change the TC of a BH spacetime. To gain further insight on this connection, in this paper we further deform the Schwarzschild-Melvin spacetime by embedding Einstein-Maxwell theory in a larger family of models: Einstein-Maxwell-\textit{dilaton} theory, with a non-minimal coupling between the dilaton and the Maxwell field, specified by a parameter $a$, $cf.$ action~\eqref{action} below. This one-parameter family of models is known to describe several special cases of interest. For:
\begin{itemize}
\item $a=0$, it describes the standard Einstein-Maxwell theory;
\item $a=1$: it describes a truncation of a low-energy effective field theory emerging from the heterotic string;
\item $a=\sqrt{3}$ it describes the original Kaluza-Klein (KK) theory, obtained by the dimensional reduction on a circle of five dimensional ($D=5$) vacuum GR.
\end{itemize}
Independently of these special cases, one may consider any value of $a$. Moreover, an appealing feature is that some well-known Einstein-Maxwell solutions ($a=0$) can be generalised to arbitrary $a$ in closed analytic form~\cite{Dowker94}. This is precisely the case of the Schwarzschild-Melvin solution, that for generic $a$ shall be dubbed Schwarzschild-dilatonic-Melvin (SdM) BH, given in Eqs.\eqref{lineel}-\eqref{sol2} below. The SdM solution can be seen as a three parameter family, described by $(M,B,a)$, although $M,B$ are integration constants and $a$ is a parameter in the action. Nonetheless, for scanning the space of solutions one can take these three parameters on equal footing. 

One of the interesting features concerning the SdM family of solutions is the status of the empty dilatonic-Melvin background ($M=0$)~\cite{Gibbons:1987ps}. Its asymptotic structure varies with $a$. Moreover, for the special KK value, $a=\sqrt{3}$, the Melvin background has a remarkably simple interpretation when KK uplifted to $D=5$: it is just flat spacetime compactified with a twist~\cite{Dowker94}. As we shall show, precisely at this value of $a$ there is a discontinuous change of the TC of the SdM family, i.e a \textit{topological transition}: for $a\leqslant \sqrt{3}$, TC$=0$, which includes the Einstein-Maxwell case studied in~\cite{MSBH21}; for $a>\sqrt{3}$, on the other hand, TC$=-1$, as for asymptotically flat BHs~\cite{PRLCunha:2020}. 

We shall split the SdM family of solutions into three distinct cases: Case I, Case IIA and Case IIB. Case I occurs for $a\leqslant \sqrt{3}$ (TC$=0$), and there may be two or zero LRs outside the horizon. Case IIA occurs for $\sqrt{3}<a<\sqrt{19/6}$ (TC$=-1$) and there may be one unstable or three (two unstable and one stable) LRs outside the horizon. Finally case IIB is described by $a \geqslant \sqrt{19/6}$ (TC=$-1$), and there is only one unstable LR outside the horizon. Using the backwards ray-tracing technique, we shall also compute the shadows and gravitational lensing of the SdM BH for each case described above, exhibiting qualitatively distinct results for each case.

This paper is organised as follows. In Section~\ref{sec2} we discuss the Einstein-Maxwell-dilaton model, describing in particular the connection to $D=5$ vacuum GR, for the KK coupling $a=\sqrt{3}$. In Section~\ref{sec3} we present the SdM solution, discuss some of its properties and show the $D=5$ interpretation of the Melvin KK solution. In Section~\ref{sec4} we discuss null geodesics in the SdM spacetime and use them as a probe of the asymptotic structure for different $a$. In Section \ref{sec5} we study the LRs in this SdM family for different values of $a$ and $B$, while in Section~\ref{sec6} we explain the results for the total number of LRs using topological arguments and show that the parameter $a$ induces a \textit{topological transition} of the TC. In Section~\ref{sec7} we discuss the shadows and lensing of a selection of SdM spacetimes. In Section~\ref{Final remarks} we present our final remarks about the TC, shadow and lensing in the SdM family of solutions. In the remaining of this paper, we shall use geometrised units such that $G=c=1$. 

\section{Einstein-Maxwell-dilaton theory}
\label{sec2}
We shall study the Einstein-Maxwell-dilaton model described by the following action:
\begin{align}
\label{action}S=\frac{1}{16\pi}\int d^4x\sqrt{-g}\left[R-2(\nabla \Phi)^2-e^{-2a\Phi}F^2 \right],
\end{align}
where $R$ is the Ricci scalar curvature computed from the metric $g_{\mu\nu}$, with determinant $g$, $\Phi$ is the dilaton field, and $F^2\equiv F_{\mu\nu}F^{\mu\nu}$ is the Maxwell invariant where $F_{\mu\nu}=\partial_\mu A_\nu-\partial_\nu A_\mu$ is the Maxwell-Faraday 2-form and $A_\mu$ the electromagnetic potential 1-form. The spacetime coordinates take values  $\mu,\nu=0,1,2,3$. 

As described in the Introduction, one may take any value for the dilatonic coupling, $a\in \mathbb{R}$. Since $a\rightarrow -a$ corresponds simply to $\Phi\rightarrow -\Phi$, we can restrict to $a\in \mathbb{R}_0^+$. There are, moreover, three special values: $a=0,1,\sqrt{3}$. The latter, in particular, provides interesting geometrical connections. For $a=\sqrt{3}$, the action~\eqref{action} is the KK dimensional reduction of $D=5$ GR without matter. Let us make this explicit, as it will be of use below. 

One considers the $D=5$ Einstein-Hilbert action 
\begin{equation}
\label{action5}S=\frac{1}{16\pi G_5}\int d^5X\sqrt{-\hat{g}}\hat{R} \ ,
\end{equation}
where $G_5$ is the five dimensional Newton's constant (which was reinserted for the sake of clarity), hatted quantities are five dimensional and $X^M=(x^\mu,y)$ are the five dimensional coordinates, with $M=(\mu,4)$ and $X^4\equiv y$ is the coordinate along the fifth dimension. Next one considers the KK ansatz; it amounts to writing the five dimensional metric $\hat{g}_{MN}$ in terms of four dimensional fields, $(g_{\mu\nu},\tilde{A}_\mu,\tilde{\Phi})$. It is assumed that the 4-dimensional fields depend on $x^\mu$ but not on $y$; the latter is assumed to be a cylindrical coordinate, $i.e$, $0\leqslant y \leqslant 2\pi \mathcal{R}$ with the endpoints identified, $y\sim y+2\pi n \mathcal{R}$, $\forall n \in \mathbb{Z}$, and $\mathcal{R}$ a fixed radius scale of the extra dimension. Thus, this assumption is called \textit{cylindrical condition}.\footnote{A possible dependence of the 4-dimensional fields on $y$ can be expressed in Fourier modes, leading to an infinite tower of, so called, KK modes.} The explicit ansatz KK is:
\begin{equation}
d\hat{s}^2=e^{\alpha \tilde{\Phi}}\left[g_{\mu\nu}dx^\mu dx^\nu + e^{2\tilde{\Phi}}(dy+\tilde{A}_\mu dx^\mu)^2 \right] \ .    
\label{kkansatz}
\end{equation}
Ignoring, for the moment, the prefactor $e^{\alpha \tilde{\Phi}}$, the KK ansatz has a simple geometrical interpretation. One describes the five dimensional spacetime as a circle fiber bundle over a 4-dimensional base spacetime with metric $g_{\mu\nu}$. At a given point of the base, the proper size of the circle is $2\pi \mathcal{R} e^{\tilde{\Phi}}$ - hence the name \textit{dilaton} for the scalar field, as it describes the variation in size of the extra dimension. Moreover, the 4-dimensional gauge potential $\tilde{A}_\mu$ describes how the fiber bundle \textit{twists}.

Expressing Eq.~\eqref{action5} in terms of the ansatz~\eqref{kkansatz}, one can integrate out the fifth coordinate due to the cylindrical condition. Then, one finds a four dimensional action which, in general, is not in the Einstein frame, except if one chooses 
\begin{equation}
    \alpha=-\frac{2}{3} \ .
\end{equation} 
With this choice, the action~\eqref{action5} becomes
\begin{align}
\label{action4}S=\frac{2\pi }{16\pi G_5}\int d^4x\sqrt{-g}\left[R-\frac{2}{3}(\nabla \tilde{\Phi})^2-\frac{1}{4}e^{2\tilde{\Phi}}\tilde{F}^2 \right] \ .
\end{align}
Rescaling $(\tilde{A}_\mu,\tilde{\Phi})$ and relating the Newton's constant in four and five dimensions as
\begin{equation}
\tilde{A}_\mu=2A_\mu \ , \quad \tilde{\Phi}=-\sqrt{3}\Phi \ , \quad \frac{G_5}{2\pi}=G_4 \ ,    
\end{equation}
where $G_4$ is the 4-dimensional Newton's constant, then Eq.~\eqref{action4} becomes Eq.~\eqref{action}, with the KK value $a=\sqrt{3}$ ($G_4$ was set to unity). 

To summarize, any solution $(g_{\mu\nu},A_\mu,\Phi)$ of Eq.~\eqref{action} with $a=\sqrt{3}$ yields a five dimensional Ricci flat metric of the form
\begin{equation}
d\hat{s}^2=e^{2\Phi/\sqrt{3}}g_{\mu\nu}dx^\mu dx^\nu + e^{-4\Phi/\sqrt{3}}(dy+2A_\mu dx^\mu)^2  \ .    
\label{kkansatz2}
\end{equation}

\section{Schwarzschild-dilatonic-Melvin solution}
\label{sec3}
A solution of the model~\eqref{action} describing a neutral BH in a Melvin-like magnetic field for arbitrary values of $a$ - the SdM solution - was found in Ref.~\cite{Dowker94}. The line element  is given by
\begin{align}
\nonumber ds^2=\Lambda(r,\theta)^\frac{2}{1+a^2}\left[-f(r)dt^2+\frac{dr^2}{f(r)}+r^2d\theta^2\right]\\
\label{lineel}+\Lambda(r,\theta)^{-\frac{2}{1+a^2}}r^2\sin^2\theta d\phi^2 \ ,
\end{align}
where the two metric functions are
\begin{align}
f(r)\equiv 1-\frac{2M}{r}, \quad \Lambda(r, \theta)\equiv 1+\left(\frac{1+a^2}{4}\right)B^2r^2\sin^2\theta \ .
\end{align}
The solution is completed by the dilaton field and the electromagnetic potential 1-form, which describes the external magnetic field. These are given by
\begin{align}
e^{-2 a \Phi}=\Lambda(r,\theta)^{\frac{2a^2}{1+a^2}}\ , \ \  A=\frac{2}{(1+a^2)B}\left[1-\frac{1}{\Lambda(r, \theta)}\right]d\phi .
\label{sol2}
\end{align}

The solution is, therefore, determined by three parameters $(M,B,a)$. Some special cases are:
\begin{itemize}
    \item for $M=0$ it reduces to a generalized Melvin solution. The original Melvin solution was found for $a=0$~\cite{Melvin:1963qx}, whereas this generalization, first found in Ref.~\cite{Gibbons:1987ps}, holds for general $a$, thus depending on both $B$ and $a$;
    \item for $B=0$,  it reduces to the Schwarzschild solution, regardless of $a$. The dilaton becomes a constant in accordance with scalar no-hair theorems - see $e.g.$ Ref.~\cite{Herdeiro:2015waa} and the solution depends solely on $M$;
    \item for $a=0$, it reduces to the Schwarzschild-Melvin  solution found by Ernst~\cite{Ernst76}, which depends on $M$ and $B$, whose TC was discussed in Ref.~\cite{MSBH21}; 
\end{itemize}

\begin{figure}
\centering
\subfigure{\includegraphics[scale=0.68]{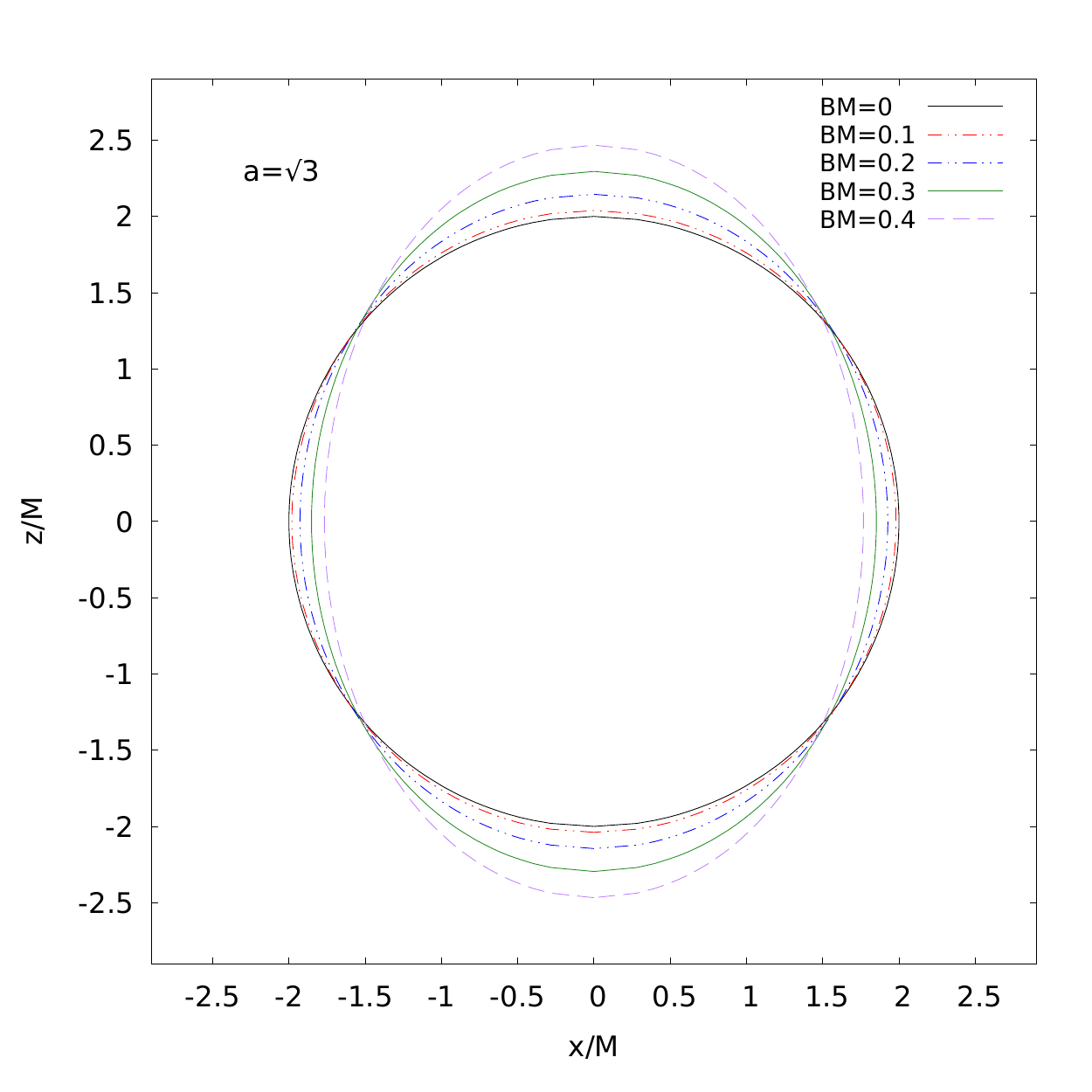}}
\subfigure{\includegraphics[scale=0.68]{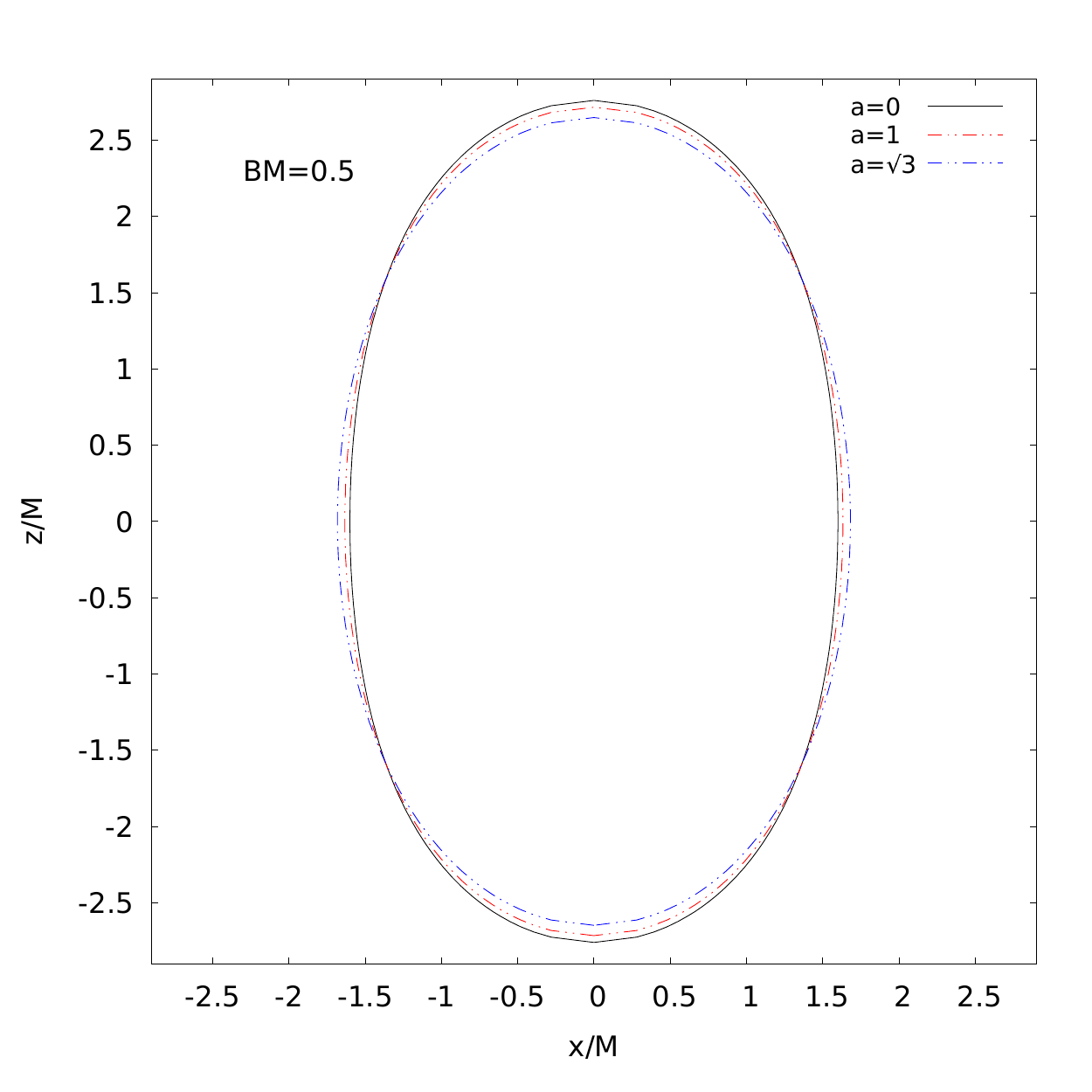}}
\caption{Embedding of the apparent horizon geometry of the SdM BH in Euclidean 3-space for: (top panel) several values of $B$ and $a=\sqrt{3}$ (KK value); (bottom panel) several values of $a$ and $BM=0.5$.} %
\label{Horizon_embedding}
\end{figure}

The line element \eqref{lineel} presents singularities at $r=0$ and $r=2M$. The former is an irremovable singularity, as can be seen by computing the Kretschmann scalar of this geometry. The latter is a coordinate singularity, and can be removed by writing the line element in, say, Eddington-Finkelstein-like coordinates. Thus one introduces the advanced time $u$,
\begin{align}
du=dt+\frac{dr}{1-\frac{2M}{r}},
\end{align}
yielding the line element
\begin{align}
\nonumber ds^2=\Lambda^{\frac{2}{1+a^2}}\left(-f(r)du^2+2dudr+r^2d\theta^2 \right)\\
+\Lambda^{-\frac{2}{1+a^2}}r^2\sin^2\theta d\phi^2,
\end{align}
which is regular at $r=2M$. In the standard Schwarzschild solution $r=2M$ corresponds to the location of the event horizon. Since the line element~\eqref{lineel} is not asymptotically flat, the definition of a teleological event horizon becomes subtle. In Ref.~\cite{MSBH21} it was shown that $r=2M$ corresponds to the location of an apparent horizon in the Schwarzschild-Melvin geometry ($a=0$). Similarly, we can also show that $r=2M$ is the location of an apparent horizon for general $a$ (see Sec. II of  Ref.~\cite{MSBH21} for further details). In Fig.~\ref{Horizon_embedding} we show the isometric embedding in Euclidean 3-space of the SdM apparent horizon for different values of $a$ and $B$. One observes that increasing $BM$ for fixed $a$ ($a$ for fixed $BM$) makes the horizon geometry more (less) prolate.

Let us close this section by considering the special case with $M=0$ and $a=\sqrt{3}$, for which the SdM solution simplifies considerably to
\begin{align}
\label{melvinKK1}
\nonumber ds^2=\sqrt{\Lambda(r,\theta)}\left[-dt^2+dr^2+r^2d\theta^2\right]
+\frac{r^2\sin^2\theta d\phi^2}{\sqrt{\Lambda(r,\theta)}} \ ,
\end{align}
where
\begin{align}
\Lambda(r, \theta)\equiv 1+B^2r^2\sin^2\theta \ 
\end{align}
and 
\begin{align}
e^{-2\Phi/\sqrt{3}}=\sqrt{\Lambda(r,\theta)} \ , \ \  A=\frac{1}{2B}\left[1-\frac{1}{\Lambda(r, \theta)}\right]d\phi \ .
\label{melvinkk2}
\end{align}
Then, using Eq.~\eqref{kkansatz2}, the corresponding five dimensional geometry is, after a slight re-arrangement:
\begin{equation}
    d\hat{s}^2=-dt^2+dr^2+r^2d\theta^2+r^2\sin^2\theta  (d\phi+Bdy)^2+ dy^2 \ .
\end{equation}
This is obviously 5-dimensional Minkowski spacetime, where the five dimensional azimuthal coordinate of the latter is
\begin{equation}
\varphi_5\equiv \phi + By \ .    
\end{equation}
Thus, the four-dimensional KK Melvin solution is simply flat five-dimensional spacetime. The four-dimensional non-triviality comes from a \textit{twisted} KK reduction. Indeed, the geometrical interpretation is that one is simultaneously identifying the cylindrical coordinate $y$ and the five-dimensional azimuthal coordinate $\varphi_5$ as
\begin{equation}
    (y,\varphi_5)\sim (y+2\pi n \mathcal{R}, \varphi_5 + 2\pi m+ 2\pi n\mathcal{R} B) \ ,
    \label{twistedid}
\end{equation}
where $m\in \mathbb{Z}$ reflects the standard periodicity of the five dimensional azimuthal coordinate. In other words, each period along $y$ induces a variation of $2\pi \mathcal{R} B$ along $\varphi_5$, where $B$ is a parameter defining this twist in $\varphi_5$. The KK reduction is then performed along the orbits of this identification, corresponding to the Killing vector $\partial_y+B\partial_{\varphi_5}$. The four-dimensional azimuthal coordinate $\phi$ is a constant along these orbits and has the standard period inherited from $\varphi_5$, $\phi\sim \phi+2\pi m$.

\section{Null geodesics and asymptotics}
\label{sec4}
\begin{figure}[h!]
\centering
\subfigure{\includegraphics[scale=0.67]{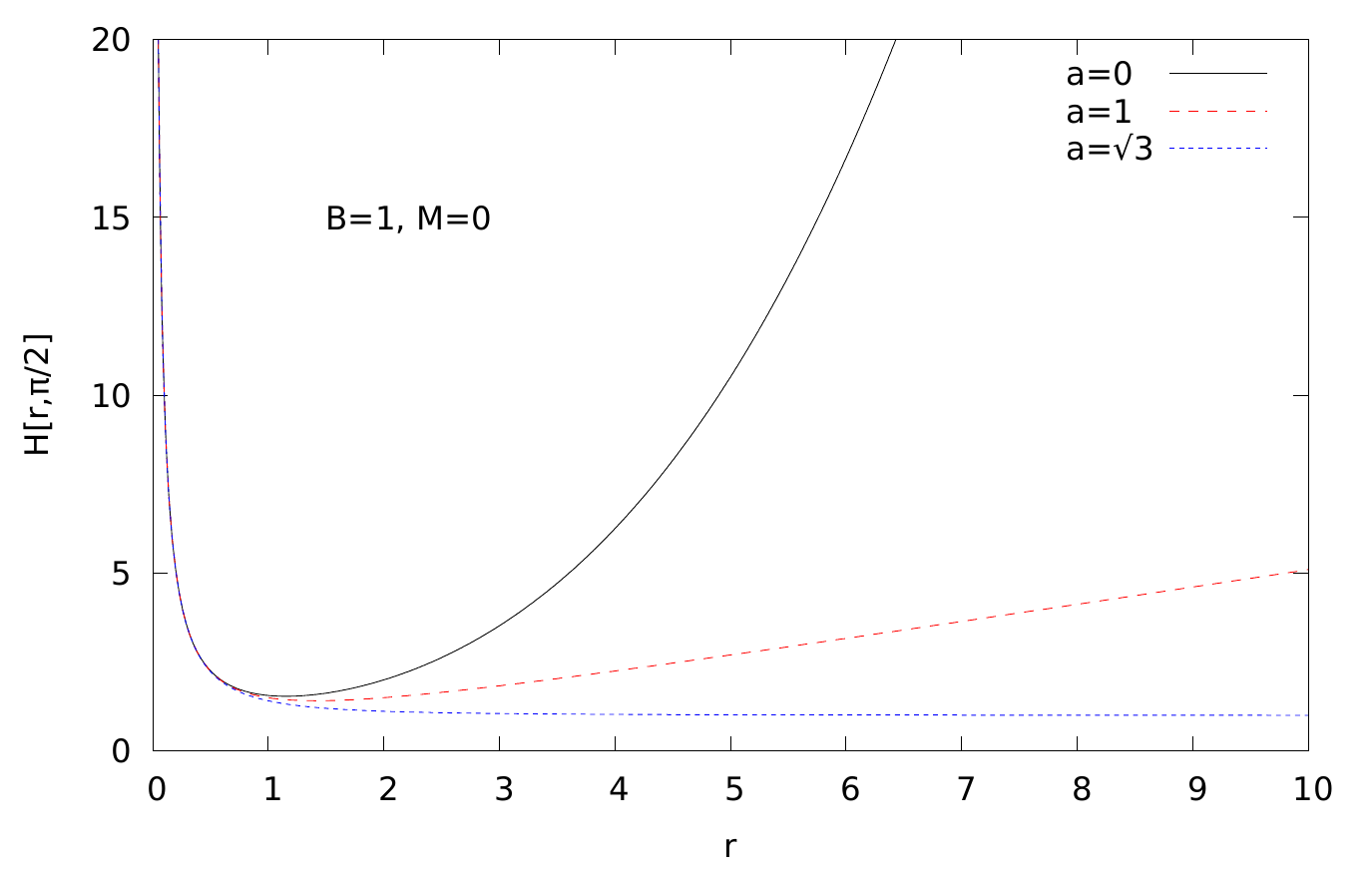}}
\subfigure{\includegraphics[scale=0.67]{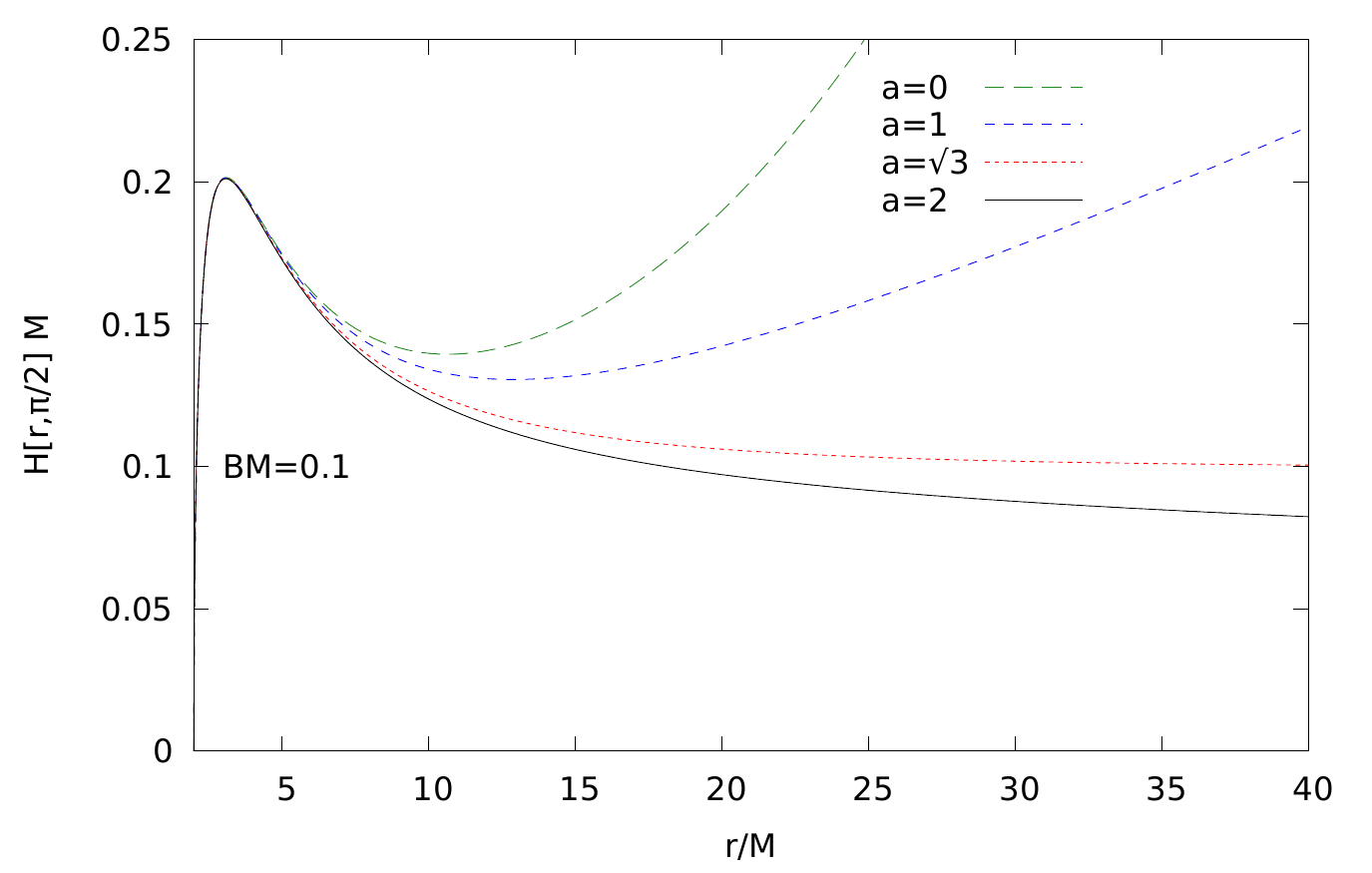}}
\subfigure{\includegraphics[scale=0.67]{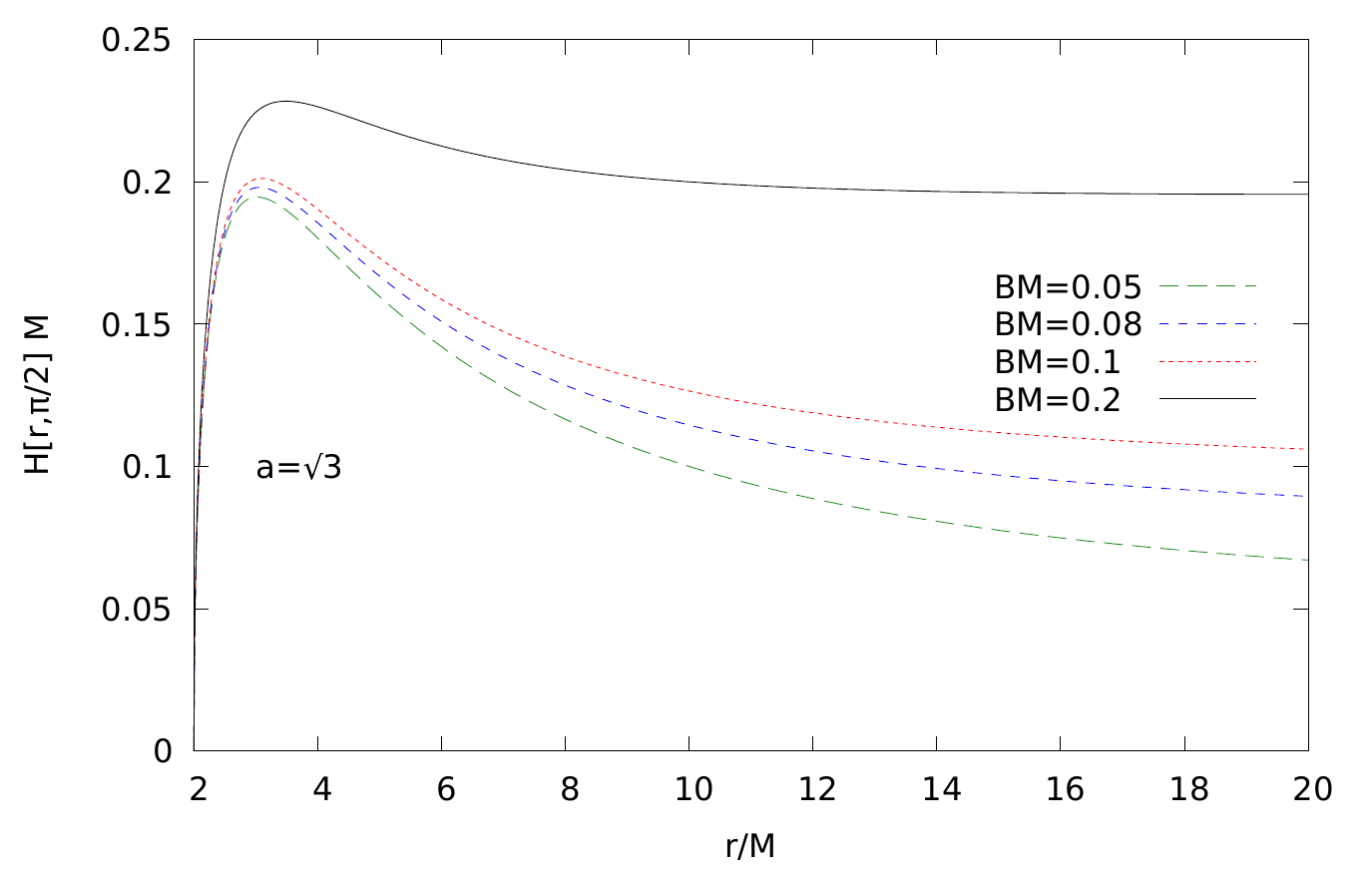}}
\caption{Effective potential $H(r,\pi/2)$ for the SdM spacetime with: (top panel) $M=0$, $B=1$ and different $a$; (middle panel) $BM=0.1$ and different $a$; (bottom panel) $a=\sqrt{3}$ and different $BM$.} %
\label{ef-potential}
\end{figure}

Let us now consider null geodesics on the SdM geometry. We adopt the Hamiltonian formalism. The Hamiltonian for null geodesics in a curved spacetime is given by
\begin{align}
\label{Hamiltonian}\mathcal{H}=\frac{1}{2}g^{\mu\nu}p_\mu p_\nu=0 \ ,
\end{align}
where $p_\mu$ is the photon's momentum 4-co-vector.
Hamilton's equations are
\begin{align}
&\dot{x}^{\mu}=\frac{\partial \mathcal{H}}{\partial p_\mu} \ ,\\
&\dot{p}_{\mu}=-\frac{\partial \mathcal{H}}{\partial x^\mu} \ ,
\end{align}
where the overdots denote differentiation with respect to the affine parameter. The spacetime's (given by Eq.~\eqref{lineel}) stationarity and axial symmetry  imply two conserved quantities along the null geodesics;  they are
\begin{align}
&p_t=-E \ ,\\
&p_\phi=L \ .
\end{align}
Taking account the constants of motion $p_t$ and $p_\phi$, the explicit form of the Hamiltonian \eqref{Hamiltonian} becomes
\begin{align}
\mathcal{H}=T(r,\theta)+V(r,\theta,E,L) \ ,
\end{align}
where
\begin{align}
&T(r,\theta)=g^{rr}(p_r)^2+g^{\theta\theta}(p_\theta)^2 \ ,\\
&V(r,\theta,E,L)=-\frac{\Lambda(r,\theta)^{-\frac{2}{1+a^2}}E}{f(r)}+\frac{\Lambda(r,\theta)^{\frac{2}{1+a^2}}L^2}{r^2\sin^2\theta} \ ,
\end{align}
are the kinetic and the potential terms, respectively. The latter can be cast in terms of a new effective potential, $H(r,\theta)$, which is independent of $E$ and $L$:
\begin{align}
V(r,\theta,E,L)=\frac{L^2 \Lambda(r,\theta)^{\frac{2}{1+a^2}}}{f(r)}\left[H(r,\theta)-\frac{1}{\eta}\right]\left[H(r,\theta)+\frac{1}{\eta}\right] 
\end{align}
and $\eta\equiv L/E$ is the impact parameter of the geodesic. For the SdM BH, this new effective potential is given by
\begin{equation}
\label{effec-pot}H(r,\theta)\equiv \frac{\Lambda(r,\theta)^{\frac{2}{1+a^2}}\,f(r)^\frac{1}{2}}{r\sin\theta} \ .
\end{equation}
This effective potential is illustrated in Fig.~\ref{ef-potential}.

The kinetic term [$T(r,\theta)$] is always non-negative. Thus, the phase space of null geodesics must obey $V(r,\theta,E,L)\leqslant 0$. Hence the motion of null geodesics obeys the following inequality:
\begin{align}
\frac{1}{|\eta|}\geqslant H(r,\theta) \ .
\end{align}

In Ref.~\cite{MSBH21} it was shown that for $a=0$ and $B\neq 0$  - the electrovacuum Melvin case - light rays with $L \neq 0$ cannot escape to infinity due to the asymptotic behavior of $H(r,\theta)$. Thus, the Einstein-Maxwell Melvin solution works as a confining box for light rays; only outwards radially directed light rays can reach $r\rightarrow \infty$. For general $a$, on the other hand,  the asymptotic behavior is modified. For the effective potential $H(r,\theta)$ we find (for $\theta\neq 0,\pi$)
\begin{align}
\label{H_infinity}\lim_{r\rightarrow \infty} H(r,\theta) \sim  r^{\frac{3-a^2}{1+a^2}} \ .
\end{align}
This implies three qualitatively distinctive behaviours. For:
\begin{itemize}
\item $a<\sqrt{3}$, light rays with $L\neq 0$ cannot escape to spatial infinity; only radially outgoing light rays reach the spatial infinity. This includes the standard Einstein-Maxwell case analysed in Ref.~\cite{MSBH21}; 
\item $a=\sqrt{3}$, light rays with $L\neq 0$ can escape to spatial infinity if $1/|\eta| \geqslant |B|$;
\item $a>\sqrt{3}$, all the light rays can escape to spatial infinity, similarly to any asymptotically flat geometry.
\end{itemize}
These behaviours can be understood from the corresponding effective potentials plotted for different values of $a$ in Fig.~\ref{ef-potential}. 
Thus, as  $a$ is increased, the SdM solution behaves as an asymptotically flat spacetime concerning the motion of null geodesics, with the transition point occurring at the KK value $a=\sqrt{3}$. This provides a context for our findings below concerning the LRs TC.

\section{Light rings}
\label{sec5}

LRs correspond to critical points of $V$. In terms of the effective potential $H(r,\theta)$, they are determined by
\begin{align}
\label{LR_Eq1}&H(r,\theta)=\frac{1}{\eta} \ ,\\
\label{LR_Eq2}&\nabla H(r,\theta)=0 \ .
\end{align}
Let us consider first the dilatonic Melvin case ($M=0$). From Eqs.~\eqref{effec-pot} and~\eqref{LR_Eq2} we find that a LR exists at $\theta=\pi/2$ and
\begin{align}
\label{MelvinLR}r=\frac{2}{\sqrt{3-a^2}B} \ .
\end{align}
For $a=0$, we recover the LR radial coordinate in the Melvin spacetime~\cite{MSBH21}. For $a\neq 0$, there are two distinct cases. For:

\begin{itemize}
\item $a< \sqrt{3}$, there is one stable LR with radial coordinate given by Eq.~\eqref{MelvinLR};
\item $a\geqslant \sqrt{3}$, there are no LRs at finite radial coordinate.
\end{itemize}
We remark that due to the additional Killing vector in the $z$ direction in the dilatonic Melvin case, there is a LR tube for $a<\sqrt{3}$ with radius given by Eq.~\eqref{MelvinLR}, instead of a single LR along the equatorial plane. As the value of $a$ approaches $\sqrt{3}$, the cylindrical radius of the LR tube approaches infinity. This result generalizes the $a=0$ case studied in Ref.~\cite{MSBH21}.

Let us now consider the SdM general case. From Eqs.~\eqref{LR_Eq1} and \eqref{LR_Eq2}, we find that LRs are located at $\theta=\pi/2$ and
\begin{align}
\label{LR_rad_coord}(3-a^2)B^2r^3+MB^2(3a^2-5)r^2-4r+12M=0 \ .
\end{align}
The number of real positive roots of Eq.~\eqref{LR_rad_coord} depends on $a$. Applying the Descartes rule of signs to Eq.~\eqref{LR_rad_coord} we obtain that: 
\begin{itemize}
\item For $a\leqslant \sqrt{3}$, we may have zero or two LRs;
\item For $a> \sqrt{3}$, we may have one or three LRs.
\end{itemize}
Again, the KK case ($a=\sqrt{3}$) provides a threshold for the total number of LRs. A confirmation on the number and location of the LRs can be done numerically. In Fig.~\ref{LR-diagram} we show the numerical results for the LRs radial coordinates, which corresponds to the roots of Eq.~\eqref{LR_rad_coord}, in terms of $B$ for fixed values of $a$. This  analysis identifies three qualitatively different cases:
 
 \begin{figure}[h!]
\includegraphics[scale=0.66]{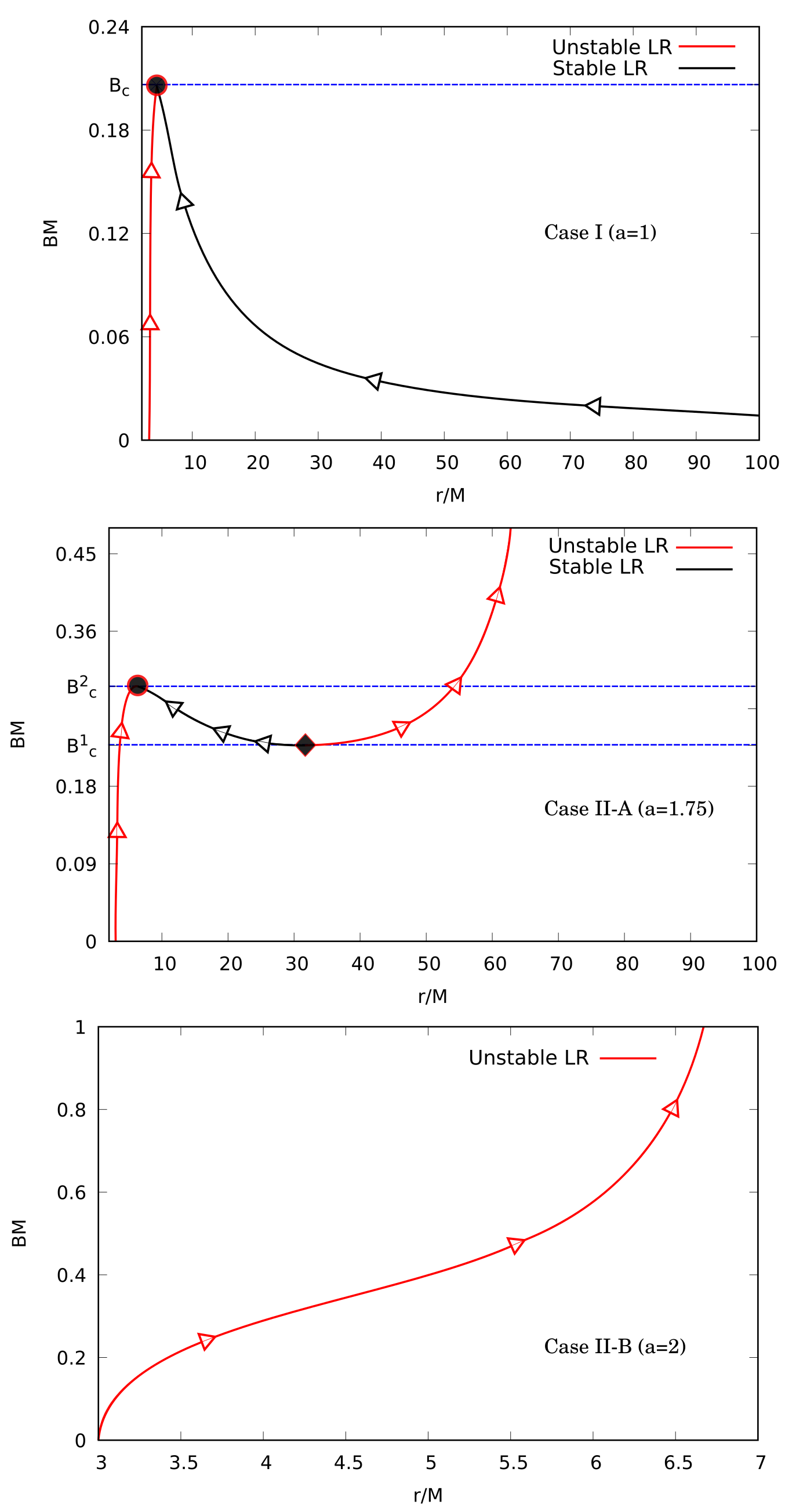}
\caption{Radial coordinate of the LRs, scanning $BM$ for illustrative cases. Top panel: case I; middle panel: case IIA; bottom panel: case IIB. The arrows indicate the direction of increasing/decreasing radial coordinate of the LRs as $BM$ increases.}
\label{LR-diagram}
\end{figure}

\begin{figure}[h!]
\centering
\includegraphics[scale=0.4]{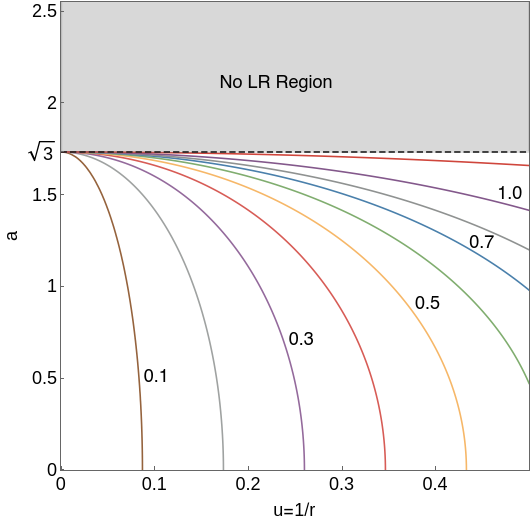}
\includegraphics[scale=0.4]{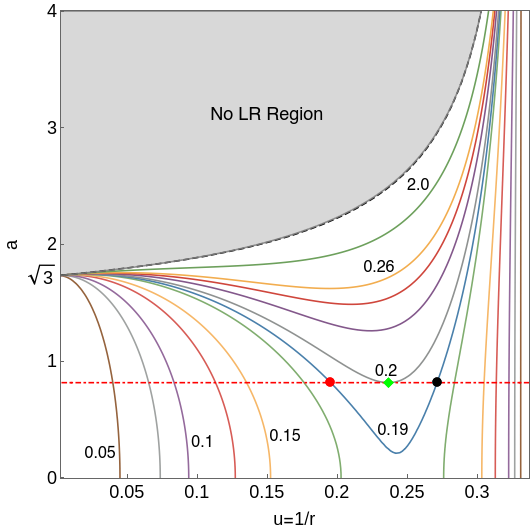}
\caption{Representation of the LRs as contour plots in the $(\text{u},a)$-plane. Each colored solid line represents a fixed value of the magnetic field $B$. We show the values of $B$ (normalized by $M$ in the SdM case) next to some colored lines. In the top panel we show the dilatonic Melvin ($M=0$) case, while in the bottom panel we show the SdM case ($M\neq 0$). The grey area denotes the region of the $(\text{u},a)$-plane where there are no LRs.
 }
\label{LR_contourplot}
\end{figure}

\begin{itemize}
\item Case I: For any $a\leqslant \sqrt{3}$, there may be zero (for $B>B_c(a)$) or two (for $B<B_c(a)$) LRs outside the horizon, where $B_c(a)$ is an $a$-dependent critical value (cf. top panel of Fig.~\ref{LR-diagram}). In the latter case, one LR is stable and one is unstable;
\item Case IIA: For any $\sqrt{3}<a<\sqrt{\frac{19}{6}}$, there may be one
(for $B<B^1_{\ c}(a)$ or $B>B^2_{\ c}(a)$) or three (for $B^1_{\ c}(a)<B<B^2_{\ c}(a)$) LRs outside the horizon, where $B^1_{\ c}(a)$ and $B^2_{\ c}(a)$ are two critical values (cf. middle panel of Fig.~\ref{LR-diagram}). In the former case the LR is unstable; in the latter case, two LRs are unstable and one is stable;
\item Case IIB: For any $a\geqslant\sqrt{\frac{19}{6}}$, there is one unstable LR outside the horizon for any value of $B$ (cf. bottom panel of Fig.~\ref{LR-diagram}).
\end{itemize}
To represent more systematically the roots of Eq.~\eqref{LR_rad_coord}, we have displayed in Fig.~\ref{LR_contourplot} the location of LRs in both dilatonic Melvin (upper panel) and SdM (lower panel) as contour plots in the $(\text{u}\equiv 1/r,a)$-plane for different values of $B$. We choose the $\text{u}$ coordinate in order to better visualize the behavior of the LRs close to spatial infinity, which is mapped to $\text{u}=0$. Each colored solid line represents a fixed value of $B$ (normalized by $M$ in the SdM case) indicated by the numbers next to some colored lines. The gray area denotes the region where there are no LRs. 
The location of the LRs in Fig.~\ref{LR_contourplot} is depicted by considering any $a=const.$ horizontal line, and then by inspecting all the intersections between that horizontal line and the colored $B=const.$ solid lines. Any such intersection yields a LR radial position for that chosen value of $a$ and for the different values of $B$.
Considering first the dilatonic Melvin case (top panel), we note that there exists at most one LR for a given value of $a$, and that its radial coordinate $r$ increases along each $B=const.$ colored line, as $a$ is increased. The black dashed line denotes the limit when $B \rightarrow \infty$ and $a\rightarrow \sqrt{3}$. 

Considering now the SdM case (bottom panel), colored solid lines are disconnected for lower values of $BM$ ($e.g.$ $BM=0.1$). As the value of $BM$ is increased, the two disconnected line pieces become connected for $BM$ larger than $\simeq 0.189$. To illustrate how this impacts on the number of LRs, we have displayed a horizontal $a=const.$ red dotted-dashed line in the bottom panel of Fig.~\ref{LR_contourplot}. For that specific horizontal red line there can be: i) two intersections with the $BM=0.19$ contour line (the red and black points), and therefore two LRs; ii) a single intersection with a $BM\simeq 0.2$ line at the green point, corresponding to a merger/formation of a stable/unstable LR pair; iii) not a single intersection with any $BM\gtrsim 0.2$ line, and thus no LRs. A similar analysis reveals that for $a>\sqrt{3}$ there are either one or three (non-degenerate) LRs. As before, the black dashed line represents the contour plot line limit $B\rightarrow \infty$. In the next section, this numerical analysis will be confirmed by computing the TC in the SdM family.

To close this section, let us consider the specific case of $a=\sqrt{3}$ and the interpretation of the corresponding LRs in $D=5$. Uplifting the SdM solution~\eqref{lineel}-\eqref{sol2} with $a=\sqrt{3}$ using the KK ansatz~\eqref{kkansatz2}, one gets the $D=5$ Ricci flat geometry
\begin{equation}
    d\hat{s}^2=-f(r)dt^2+\frac{dr^2}{f(r)}+r^2d\theta^2+r^2\sin^2\theta  (d\phi+Bdy)^2+ dy^2 \ .
\end{equation}
This is simply the four-dimensional Schwarzschild solution with one added flat direction, $y$. The radial equation for null geodesics with $\theta=\pi/2$ in this geometry is $\dot{r}^2=E^2-V_{\rm eff}$, where the effective potential is
\begin{equation}
V_{\rm eff}=f(r)\left[\frac{L_5}{r^2}+p_y^2\right] \ . 
\label{efeV}
\end{equation}
Here, $L_5$ and $p_y$ are the conserved angular momentum and momentum associated to the $D=5$ Killing vector fields $\partial_{\varphi_5}$ and $\partial_y$, respectively. The critical points of Eq.~\eqref{efeV}, obeying both $V'_{\rm eff}=0$ and $V_{\rm eff}=E^2$, determine the LRs. Solving these equations leads to a family of LRs, with the values of $\{r,L_5\}$ parameterized by $|p_y| \in [0,\sqrt{9/8}E]$.

For instance, if one imposes no motion in $y$, $p_y=0$, then there is only one critical point at $r=3M$, as expected from the Schwarzschild solution, as the extra dimension does not contribute for the motion in this case. More generically, the twisted identification~\eqref{twistedid}, that leads to the $D=4$ KK SdM, imposes
\begin{equation}
p_y=BL_5 \ .    
\end{equation}
It follows that the critical points of Eq.~\eqref{efeV} obey
\begin{equation}
MB^2r^2-r+3M=0 \ ,    
\end{equation}
which is precisely Eq.~\eqref{LR_rad_coord} for $a=\sqrt{3}$. Thus, in the KK case, the SdM LRs correspond, in $D=5$, to LR orbits moving both in $\varphi_5$ and $y$, with the corresponding momenta related by the twisted identification.

\section{{Topological charge}}
\label{sec6}
The analysis of the TC in stationary, axi-symmetric and
asymptotically flat BH spacetimes was discussed in Ref.~\cite{PRLCunha:2020}. Since the SdM spacetime is not asymptotically flat, that analysis needs to be adapted, following Ref.~\cite{MSBH21}. 

Let us introduce the vector field $\textbf{v}=(\text{v}_r,\text{v}_\theta)$, where the components are given by
\begin{align}
\text{v}_i \equiv \frac{\partial_i H}{\sqrt{g_{ii}}} \ , \quad i=(r,\theta) \ .
\end{align}
Hence the components of the vector field $\textbf{v}$ for the SdM BH are
\begin{align}
\nonumber &\text{v}_r=\frac{\sin\theta}{4r^3\Lambda^\frac{a^2}{1+a^2}}\left[\left(3-a^2\right)B^2r^3 + \left(3a^2-5\right)MB^2r^2 \right.\\
\label{vr}&\qquad \qquad \qquad \left. -4\left(r-3\,M \right)\csc^2\theta \right] \ ,\\
\label{vtheta}&\text{v}_\theta=\frac{\sqrt{1-\frac{2M}{r}}\,\cos\theta}{4r^2\Lambda^\frac{a^2}{1+a^2}}\left[\left(3-a^2\right)B^2r^2-4\csc^2\theta \right] \ .
\end{align}

\begin{widetext}

\begin{figure}[h!]
\centering
{\includegraphics[scale=0.66]{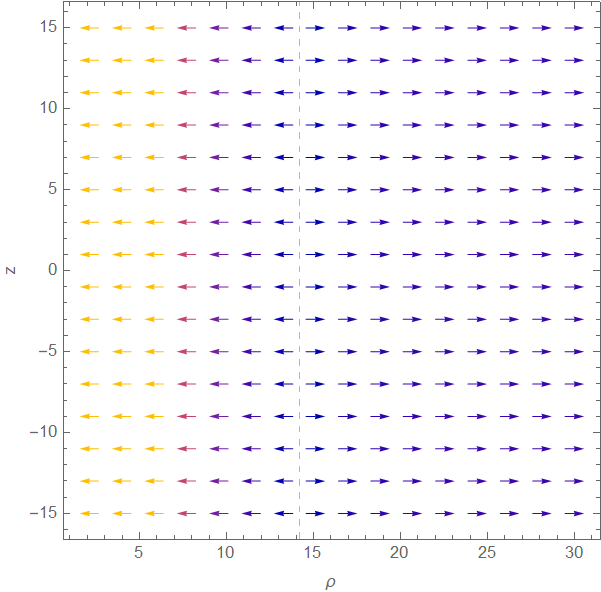}}
{\includegraphics[scale=0.66]{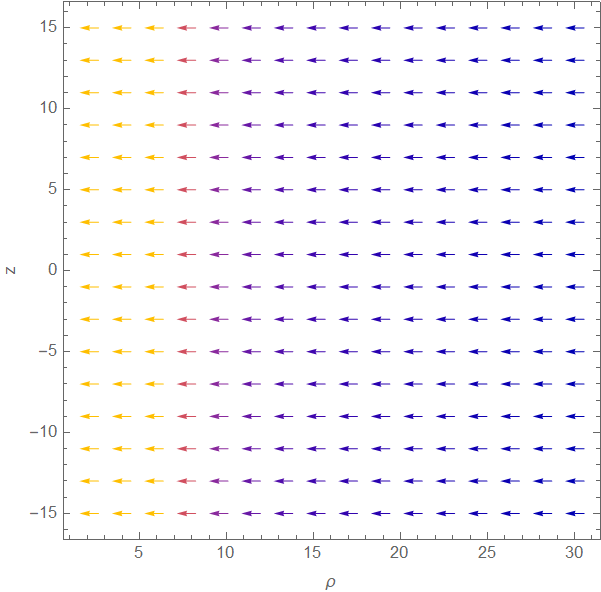}}\\
{\includegraphics[scale=0.39]{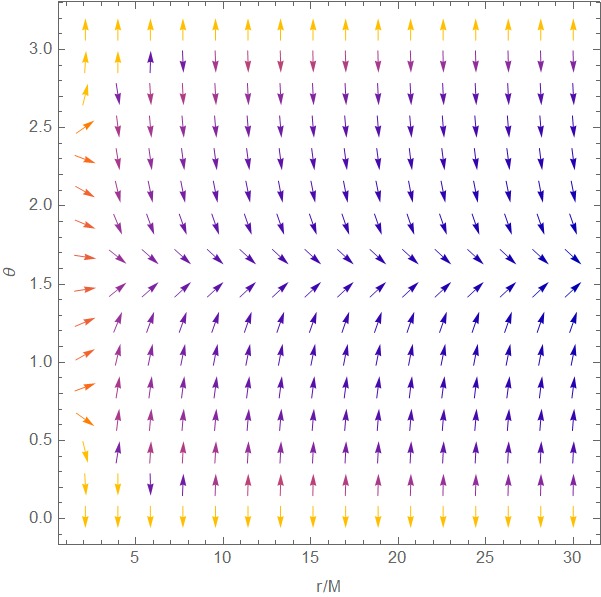}}
{\includegraphics[scale=0.39]{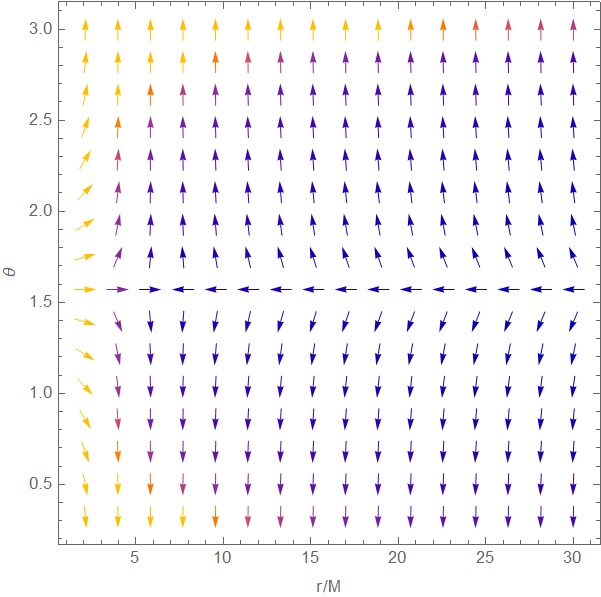}}
\caption{Top row: Plot of the normalized vector field $\textbf{v}$ in the ($\rho$, $z$)-plane, for the dilatonic Melvin spacetime with $a=1$ and $B=1$ (left panel) and $a=\sqrt{3}$ (right panel). Bottom panel: Plot of the normalized vector field $\textbf{v}$ in the ($r$,~$\theta$)-plane, for the SdM spacetime with $a=1$ (left panel) and $a=2$ (right panel). The red line in the top row left panel denotes the location of the stable LR tube.  In this figure we have chosen $B=1$ for the dilatonic Melvin spacetime and $BM=1$ for the SdM spacetime.}
\label{vector_fields}
\end{figure} 
\end{widetext}

We can write these components of the vector field in terms of a norm $v$ and the angle $\Omega$, as given by
\begin{align}
&\text{v}_r=\text{v}\cos\Omega \ ,\\
\label{vtheta_comp}&\text{v}_\theta=\text{v}\sin\Omega \ ,
\end{align}
where the norm $v$ is written as
\begin{align}
\text{v}^2=\text{v}_r^2+\text{v}_\theta^2 \ .
\end{align}
In the top row of Fig.~\ref{vector_fields}, we represent the vector field $\textbf{v}$ in the $(\rho,z)$ space for the dilatonic Melvin with $a=1$ (left panel) and $a=\sqrt{3}$ (right panel). In the bottom row of Fig.~\ref{vector_fields} we represent the vector field $\textbf{v}$ in the $(r,\theta)$ space for the SdM spacetime with $a=1$ (left panel) and $a=2$ (right panel). We notice that the parameter $a$ modifies the asymptotic behavior of the vector field $\textbf{v}$.

 
 We compute the total TC ($w$) by considering a piecewise
smooth and positive oriented curve $\mathcal{C}$ in the ($r,\theta$) space. The total TC is given by the following integer quantity
\begin{align}
\label{w_contour}w=\frac{1}{2\pi}\oint_{\mathcal{C}}d\Omega \ .
\end{align}
One can deform the contour $\mathcal{C}$, without intersecting the LRs, and the value of $w$ remains unaltered. A standard unstable LR contributes with $w=-1$ to the total TC, while a standard stable LR contributes with $w=+1$~\cite{Cunha:2017qtt,PRLCunha:2020}. For any value of $a$ and $MB$, we choose a finite contour $\mathcal{C}$ such that all the LRs outside the event horizon are placed inside the contour. We choose the same contour adopted in Ref.~\cite{PRLCunha:2020} (see, for instance, Fig.~2 of Ref.~\cite{PRLCunha:2020}), where $\mathcal{C}=L_1 \cup L_2 \cup L_3 \cup L_4$. Thus we can separate the integral \eqref{w_contour} into four distinct integrals:
\begin{align}
\label{w_int}2\pi w_{\mathcal{I}}=\mathcal{I}_1+\mathcal{I}_2+\mathcal{I}_3+\mathcal{I}_4,
\end{align} 
where
\begin{align}
&\mathcal{I}_1=\left[\int_\epsilon^{\pi-\epsilon} \frac{d\Omega}{d\theta}d\theta \right]_{r=R},\\
&\mathcal{I}_2=\left[\int^{r_0}_{R} \frac{d\Omega}{dr}dr \right]_{\theta=\pi-\epsilon},\\
&\mathcal{I}_3=\left[\int^\epsilon_{\pi-\epsilon} \frac{d\Omega}{d\theta}d\theta \right]_{r=r_0},\\
&\mathcal{I}_4=\left[\int_{r_0}^{R} \frac{d\Omega}{dr}dr \right]_{\theta=\epsilon}.
\end{align}
The TC outside the BH apparent horizon is obtained by considering the limit
\begin{align}
\label{w_lim}w=\lim_{R\rightarrow \infty}\lim_{r_0\rightarrow 2M}\left(\lim_{\epsilon \rightarrow 0}w_{\mathcal{I}}\right) \ .
\end{align}
In what follows we discuss in details each limit in Eq.~\eqref{w_int} for the SdM BH.
\\

\begin{itemize}
\item \textbf{SdM axis limit:}
\end{itemize}
Let us first analyze the behavior of the field $\text{v}$ in the SdM metric when we approach the axis of symmetry, $i.e.$, for $\theta\rightarrow 0,\pi$. From Eq.~\eqref{vtheta_comp} we have that
\begin{align}
\label{Omega_axis}\Omega=\arcsin\left(\frac{v_\theta}{v}\right).
\end{align}
Using Eqs.~\eqref{vr} and \eqref{vtheta}, we can compute the value of $\Omega$, given in Eq.~\eqref{Omega_axis}, as one approaches the axis of symmetry:
\begin{align}
\left. \Omega \right|_{\theta=0,\pi}=\begin{cases}
-\frac{\pi}{2}, \quad &\text{for\ } \theta \rightarrow 0,\\
+\frac{\pi}{2}, \quad &\text{for\ } \theta \rightarrow \pi.
\end{cases}
\end{align}
The conclusion drawn from the above result is that the vector $\text{v}$ becomes constant as one approaches the symmetry axis. The vector $\text{v}$ points upwards (downwards) at $\theta=\pi$ ($\theta=0$) [see bottom row in Fig.~\ref{vector_fields}]. Hence the integrals $\mathcal{I}_2$ and $\mathcal{I}_4$ do not contribute to $w$, since they approach zero as $\epsilon \rightarrow 0$. This result was first obtained in Ref.~\cite{PRLCunha:2020} for asymptotically flat BHs and it is also valid for the SdM case. This can be explained by the fact that along the axis of symmetry, the line element~\eqref{lineel} is equal to the Schwarzschild geometry. Therefore the axis limit behavior is similar to that of an asymptotically flat geometry.

\begin{itemize}
\item \textbf{SdM horizon limit:}
\end{itemize}
For the horizon limit of the SdM geometry, it is sufficient for our purpose to analyze only the sign of $\text{v}_r$ as $r_0 \rightarrow r_h=2M$. From Eq.~\eqref{vr}, we obtain that in the horizon limit:
\begin{align}
\label{vr_horizon}\left.\text{v}_r\right|_{r_h}=\frac{1}{4M\sqrt{\left. g_{\phi\phi}\right|_{r_h}}} \ .
\end{align}
Therefore we conclude that the radial component of $\text{v}$ is positive in the horizon limit, regardless of the value of $a$. Hence, along $\mathcal{I}_3$, $\text{v}$ winds in the clockwise direction, as we integrate along $\mathcal{I}_3$ in the counter-clockwise direction, pointing upwards (downwards) at $\theta=\pi$ ($\theta=0$), and resulting in
\begin{align}
\mathcal{I}_3=\Omega^{r_h}_{\theta=0}-\Omega^{r_h}_{\theta=\pi}=-\pi \ .
\end{align}
Before proceeding to the asymptotic limit, let us remark that in Ref.~\cite{PRLCunha:2020}, the following general expression was obtained for the horizon limit
\begin{align}
\label{vr_horizon_PRL}\left. \text{v}_r \right|_{r_h}=\frac{\kappa}{\sqrt{\left.g_{\phi\phi}\right|_{r_h}}} \ ,
\end{align}
where $\kappa$ is the surface gravity along the horizon. Comparing Eqs.~\eqref{vr_horizon} and \eqref{vr_horizon_PRL} we obtain that the surface gravity for the SdM BH is given by
\begin{align}
\kappa=\frac{1}{4M} \ ,
\end{align}
which is equal to the Schwarzschild case.  This result implies, for instance, that the Hawking temperature of the SdM and Schwarzschild BHs are equal, which is in accordance with Ref.~\cite{Radu:2002}, where it was shown that all the thermodynamic properties of a Schwarzschild BH are unaffected by the external magnetic field.

\begin{itemize}
\item \textbf{SdM asymptotic limit}
\end{itemize}
As we discussed before, the axis and horizon limits are similar to that of an asymptotically flat BHs. This is not the case for the asymptotic limit of the SdM BH, as we show now. The expansion of the radial component of the vector $\text{v}$ as $R\rightarrow \infty$ is given by
\begin{align}
\nonumber&\left. v_r \right|_{\infty}\approx \frac{\left(3-a^2\right)\Lambda^{\frac{1}{1+a^2}}}{\left(1+a^2\right)r^2\sin\theta}+\frac{\left(3a^2-5 \right)M\Lambda^\frac{1}{1+a^2}}{\left(1+a^2\right)r^3\sin\theta}\\
\label{vr_infinity}&\qquad\quad +\mathcal{O}\left(\frac{1}{r^\frac{1+2a^2}{1+a^2}}\right) \ .
\end{align}
From Eq.~\eqref{vr_infinity} we note that the asymptotic limit analysis must be separated in three different branches: $a<\sqrt{3}$, $a=\sqrt{3}$, and $a>\sqrt{3}$.

For $a<\sqrt{3}$, the leading term in Eq.~\eqref{vr_infinity} is positive. This means that along $\mathcal{I}_1$, $\text{v}$ winds in the counter-clockwise direction as we integrate along $\mathcal{I}_1$ in the counter-clockwise direction, resulting in
\begin{align}
\mathcal{I}_1=\Omega^{\infty}_{\theta=0}-\Omega^{\infty}_{\theta=\pi}=\pi, \quad \text{if} \quad a<\sqrt{3} \ .
\end{align}

For $a=\sqrt{3}$, the first term on the right hand side of Eq.~\eqref{vr_infinity} vanishes, and the second term remains greater than zero. Hence the leading order in this case is positive and we obtain again that
\begin{align}
\mathcal{I}_1=\Omega^{\infty}_{\theta=0}-\Omega^{\infty}_{\theta=\pi}=\pi, \quad \text{if} \quad a=\sqrt{3} \ .
\end{align}

For $a>\sqrt{3}$, the first term in the right hand side of Eq.~\eqref{vr_infinity} becomes negative, and the picture changes completely in comparison to the other cases. For $a>\sqrt{3}$, $\text{v}$ winds in the clockwise direction as we integrate along $\mathcal{I}_1$ in the counter-clockwise direction, resulting in
\begin{align}
\mathcal{I}_1=\Omega^{\infty}_{\theta=0}-\Omega^{\infty}_{\theta=\pi}=-\pi, \quad \text{if} \quad a>\sqrt{3} \ .
\end{align}
This result is similar to the one obtained for an asymptotically flat BH geometry in Ref.~\cite{PRLCunha:2020}.

\begin{itemize}
\item \textbf{SdM total TC:}
\end{itemize}
Summarizing the results obtained above for the four different paths $\lbrace \mathcal{I}_1, \mathcal{I}_2, \mathcal{I}_3, \mathcal{I}_4 \rbrace$, we obtain the following total TC for SdM:
\begin{align}
\label{TC_SdM}w=\begin{cases}
0, \quad &\text{if} \quad a\leqslant \sqrt{3} \ ,\\
-1, \quad &\text{if} \quad a> \sqrt{3} \ .
\end{cases}
\end{align}
The result \eqref{TC_SdM} explains, through topological arguments, the existence of the two different cases (case I and case II) studied in Sec.~\ref{sec5}, since a \textit{topological transition} happens exactly at the KK case ($a=\sqrt{3}$). Let us remark that this topological transition is only possible due to the role played by the parameter $a$ in the asymptotic limit. For $a>\sqrt{3}$, the SdM geometry behaves as an asymptotically flat BH concerning the motion of null geodesics.

\textbf{dilaton Melvin TC:}
Finally, let us remark that the contour integral approach to compute the TC cannot be applied to the dilatonic Melvin case ($M=0$) with $a\leqslant \sqrt{3}$, due to the existence of the stable LR tube for $a<\sqrt{3}$, and the stable LR at spatial infinity for $a=\sqrt{3}$. In both cases, the contour $\mathcal{C}$ would intersect at least one LR, hence a different approach is required.

For $a>\sqrt{3}$ (and $M=0$), however, we can adopt a contour $\mathcal{C}$ without intersecting any LR, and then compute the TC. The results for the axis limit and the asymptotic limit will be exactly the same as in the SdM case. The main difference arises due to the absence of an event horizon, since the vector field in the radial direction behaves as
\begin{align}
\left. v_r \right|_{r= 0} \approx -\frac{1}{r^2\sin\theta}+\frac{3}{4}B^2\sin\theta+\mathcal{O}\left(r^2\right),
\end{align}
next to the origin. The radial component of the vector field $\text{v}$ is negative in the dilaton Melvin case, in contrast to the SdM case with $a> \sqrt{3}$. This implies that the TC for the dilaton Melvin case is equal to
\begin{align}
w=0, \quad \text{if}\quad a>\sqrt{3}\quad  \text{and}\quad  M=0,
\end{align}
which explains why the dilaton Melvin solution with $a> \sqrt{3}$ has no LRs at all for $r>0$.

\section{Shadows and gravitational lensing}
\label{sec7}

\subsection{Observer setup}
We now study the gravitational lensing and shadows in SdM and dilatonic Melvin spacetimes. We apply the so-called backwards ray-tracing method in order produce a computational simulation of the optical appearance of these geometries. We solve numerically the following equations of motion:
\begin{align}
\label{dott}&\dot{t}=\frac{E}{\Lambda^\frac{2}{1+a^2}\,\left(1-\frac{2M}{r}\right)},\\
&\dot{\phi}=\frac{L\Lambda^\frac{2}{1+a^2}}{r^2\sin^2\theta},\\
&\ddot{r}+\Gamma^{r}_{\ \mu\nu}\dot{x}^\mu\dot{x}^\nu=0,\\
\label{dottheta}&\ddot{\theta}+\Gamma^{\theta}_{\ \mu\nu}\dot{x}^\mu\dot{x}^\nu=0,
\end{align}
where $\Gamma^{\alpha}_{\mu\nu}$ are the Christoffel symbols for the  geometry. The initial conditions for the system are obtained computing the 4-momentum of the photon in the \textit{vierbein} of a static observer:

\begin{align}
&\hat{\lambda}^{\hat{0}}_{\ \mu}=\left(\sqrt{1-\frac{2M}{r}}\Lambda^\frac{1}{1+a^2},\ 0,\ 0,\ 0 \right),\\
&\hat{\lambda}^{\hat{1}}_{\ \mu}=\left(0, \frac{\Lambda^{\frac{1}{1+a^2}}}{\sqrt{1-\frac{2M}{r}}},\ 0,\ 0\right),\\
&\hat{\lambda}^{\hat{2}}_{\ \mu}=\left(0\ ,\ 0,\ r \Lambda^{\frac{1}{1+a^2}},\ 0 \right),\\
&\hat{\lambda}^{\hat{3}}_{\ \mu}=\left(0,\ 0,\ 0,\ \frac{r\sin\theta}{\Lambda^\frac{1}{1+a^2}} \right).
\end{align}
The components of the photon 4-momentum projected in the \textit{vierbein} ($p^{\hat{a}}=\hat{\lambda}^{\hat{a}}_{\ \mu}\,p^\mu$) are given by 
\begin{align}
\label{4momentum_1}&p^{\hat{t}}=\frac{E}{\sqrt{1-\frac{2M}{r}}\Lambda^{\frac{1}{1+a^2}}},
\qquad p^{\hat{r}}=\frac{\Lambda^\frac{1}{1+a^2}}{\sqrt{1-\frac{2M}{r}}}\dot{r},\\
\label{4momentum_2}&p^{\hat{\theta}}=r\Lambda^\frac{1}{1+a^2}\dot{\theta},
\qquad p^{\hat{\phi}}=\frac{L\Lambda^{\frac{1}{1+a^2}}}{r\sin\theta}.
\end{align}
We can parametrize the photon linear 3-momentum $\textbf{p}$ in terms of the angles  $(\alpha, \beta)$ as given by
\begin{align}
\label{pr}&p^{\hat{r}}=|\textbf{p}|\cos\alpha\cos\beta,\\
\label{ptheta}&p^{\hat{\theta}}=|\textbf{p}|\sin\alpha,\\
\label{pphi}&p^{\hat{\phi}}=|\textbf{p}|\cos\alpha\sin\beta,
\end{align}
and from Eqs.~\eqref{4momentum_1}-\eqref{ptheta}, we obtain that
\begin{align}
&E=|\textbf{p}|\sqrt{1-\frac{2M}{r}}\Lambda^{\frac{1}{1+a^2}},\\
&\dot{r}=|\textbf{p}|\frac{\sqrt{1-\frac{2M}{r}}}{\Lambda^\frac{1}{1+a^2}}\cos\alpha\cos\beta,\\
&\dot{\theta}=\frac{|\textbf{p}|}{r\Lambda^\frac{1}{1+a^2}}\sin\alpha,\\
&L=\frac{|\textbf{p}|\,r\,\sin\theta}{\Lambda^\frac{1}{1+a^2}}\cos\alpha\sin\beta,
\end{align}
which are the initial conditions for Eqs.~\eqref{dott}-\eqref{dottheta}. We solve the equations of motion from the observer position, and backwards in time, until the light ray is scattered to infinity or absorbed by the event horizon (when present). Each pair $(\alpha, \beta)$ represents a point in the image plane of the observer. If the light ray, evolved backwards in time, hits the event horizon, we assign a black color to the coordinate $(\alpha, \beta)$ in the image. On the other hand, if the light ray is scattered to a celestial sphere of radius $r_{cs}$, we assign a color to the coordinate $(\alpha, \beta)$. We use the same pattern of colors as in Refs.~\cite{Bohn:2015,PRLCunha:2015,GRGCunha:2018}, where the celestial sphere is divided in four quadrants (red, green, blue, yellow).

\begin{figure}[h!]
\includegraphics[scale=0.66]{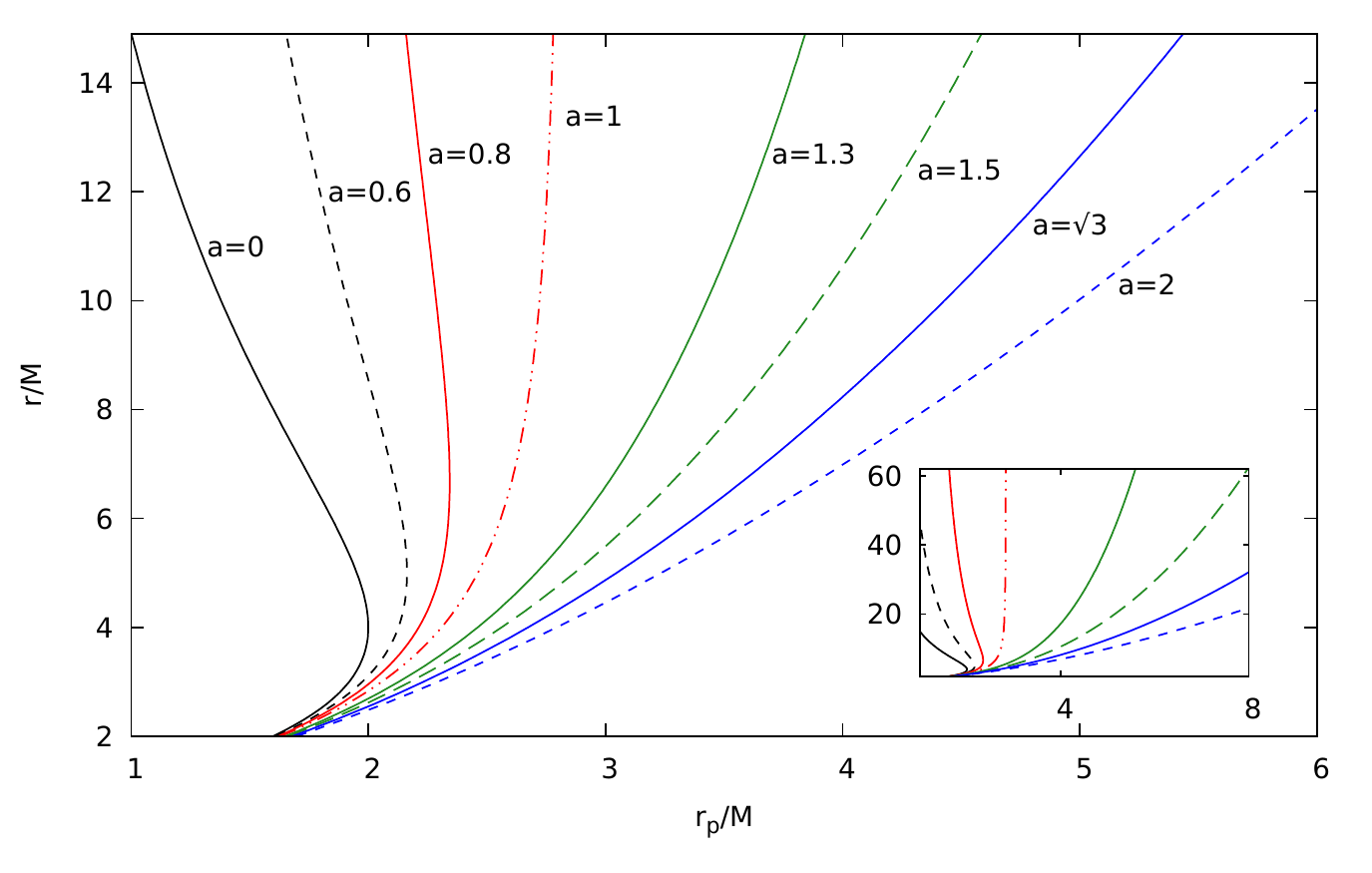}
\caption{Behavior of the perimetral radius as a function of $r$ for several values of $a$, and $BM=0.5$. Observe that for $a=1$ the perimetral radius tends to a constant value at spatial infinity, while for $a>1$ it is a monotonically increasing function of the radial coordinate.}
\label{perimetral_radius}
\end{figure}

In order to place the observer in similar observational condition, we adopt the perimetral radius ($r_p$) as a geometrically invariant measure of distance:
\begin{align}
r_p\equiv \left.\sqrt{g_{\phi\phi}} \right|_{\theta=\pi/2}=\frac{r}{\left[1+\frac{\left(1+a^2\right)}{4}B^2\,r^2 \right]^\frac{1}{1+a^2}} \ .
\end{align}
We note that the perimetral radius has a maximum value at
\begin{align}
\left. r_p \right|_{\text{max}}=\frac{2}{B\sqrt{1-a^2}} \ , \quad 0\leqslant a<1 \ .
\end{align}
In Fig.~\ref{perimetral_radius}, the perimetral radius is compared with the coordinate radius $r$ for several values of $a$, and $BM=0.5$. For $a=1$, the perimetral radius tends to a constant value at the spatial infinity, while for $a>1$ it is a monotonically increasing function of the radial coordinate $r$, as can be seen in Fig.~\ref{perimetral_radius}. It is remarkable that this transition in the behaviour of the perimetral radius occurs at another special value of $a$, as emphasised in the Introduction.

\subsection{Gravitational lensing for the dilatonic Melvin spacetime}
\begin{figure}
\centering
\subfigure[$a=0$]{\includegraphics[scale=0.12]{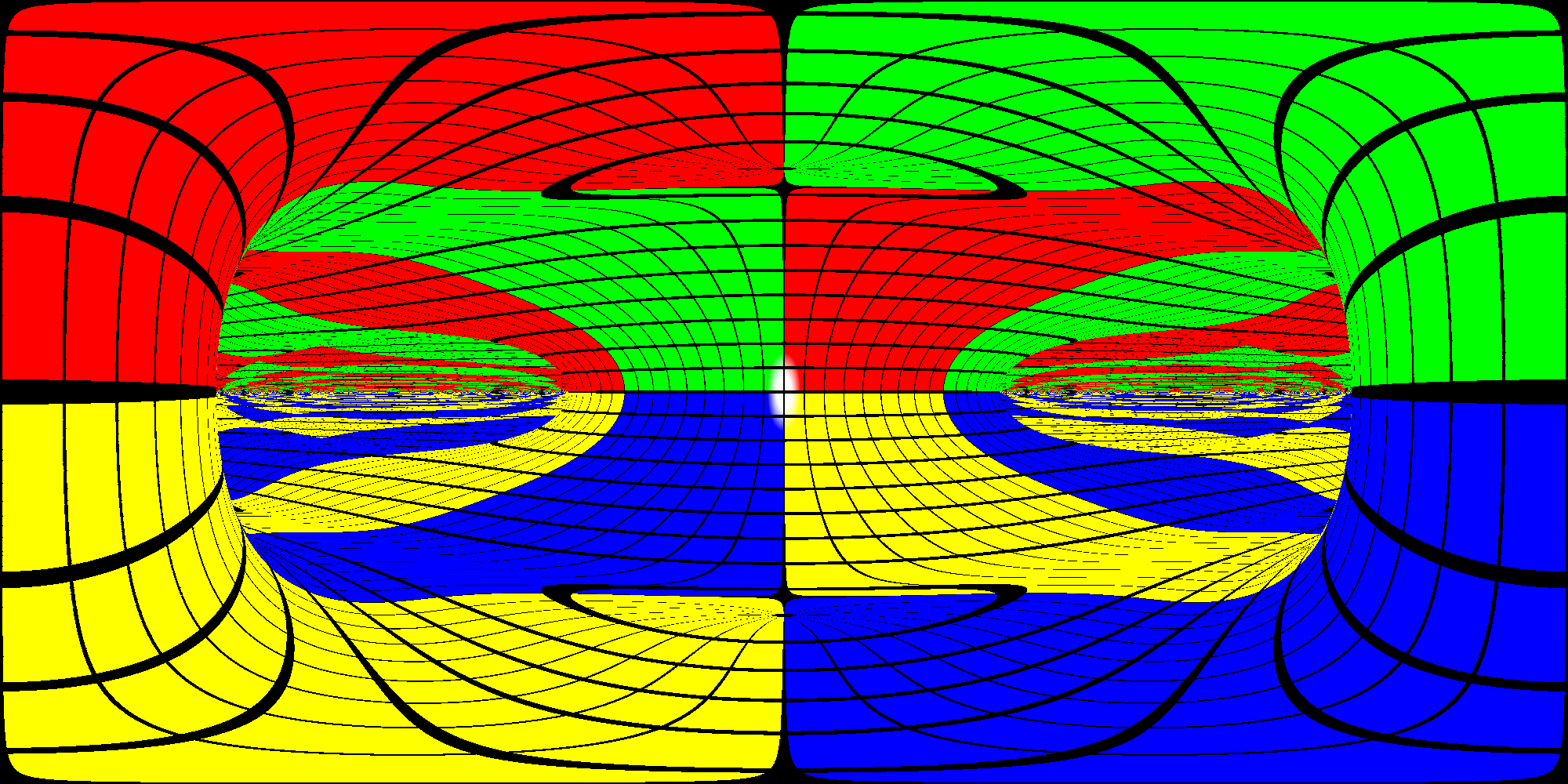}}
\subfigure[$a=1$]{\includegraphics[scale=0.12]{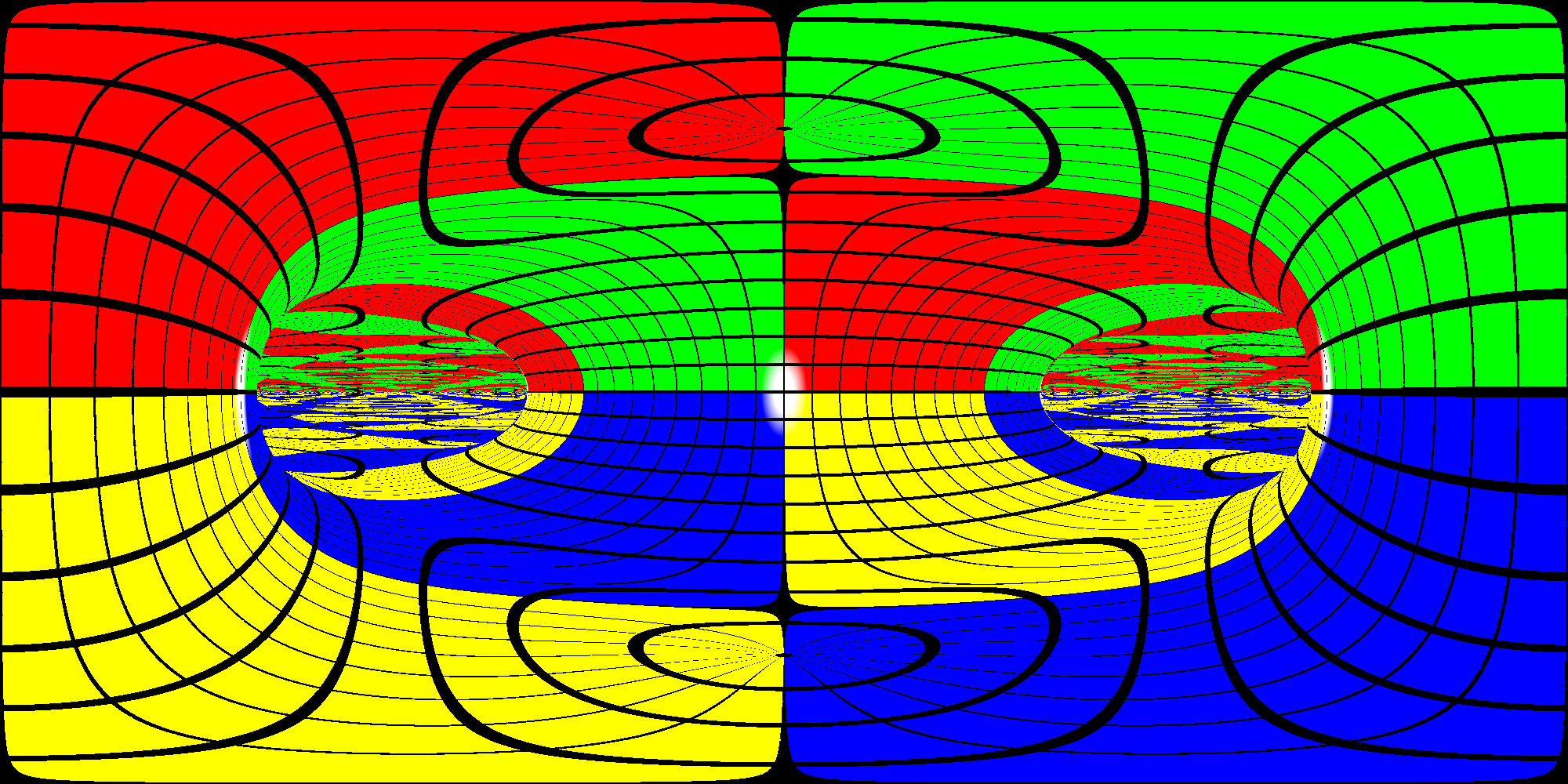}}
\subfigure[$a=\sqrt{3}$]{\includegraphics[scale=0.12]{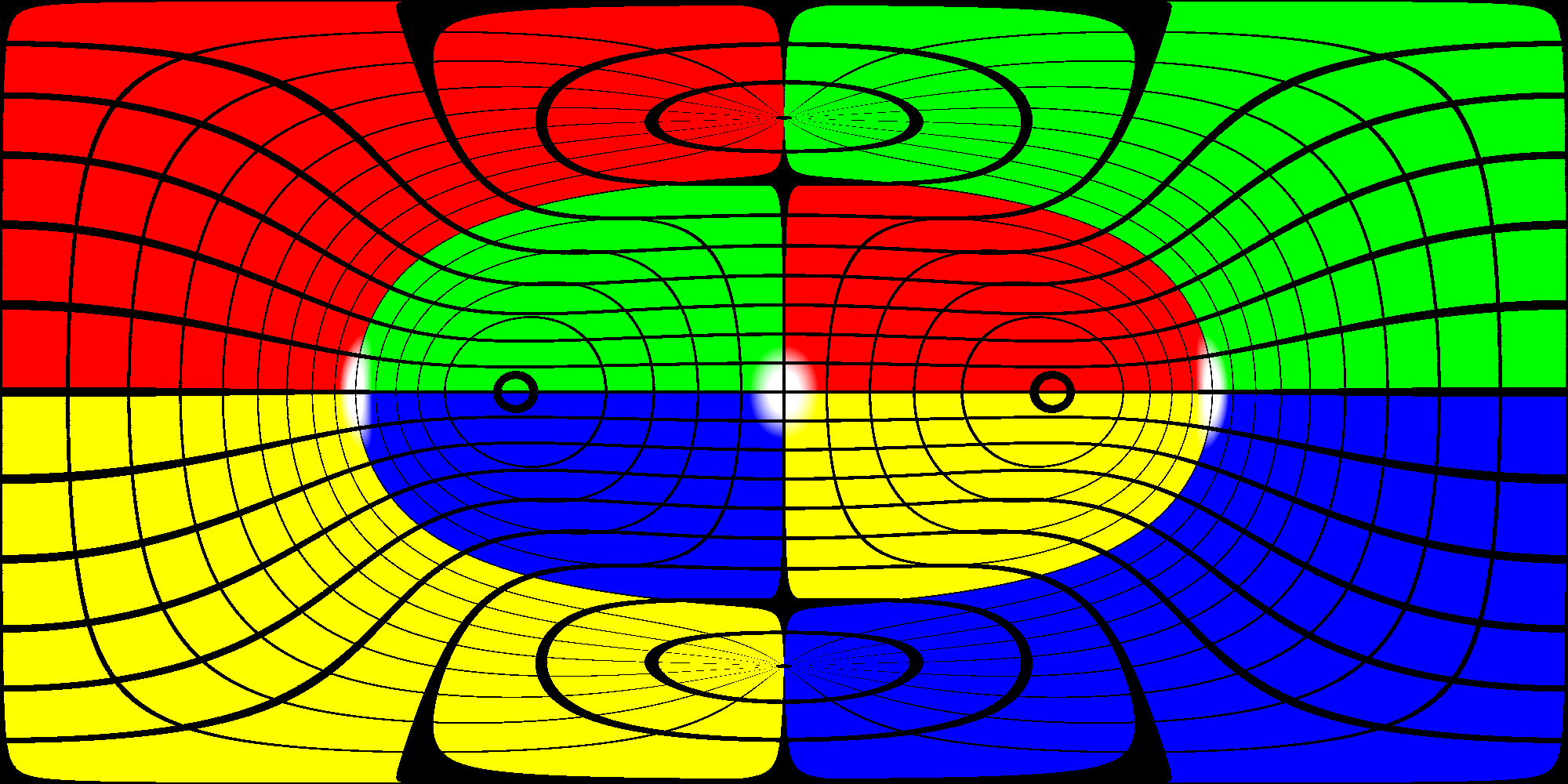}}
\subfigure[$a=2$]{\includegraphics[scale=0.12]{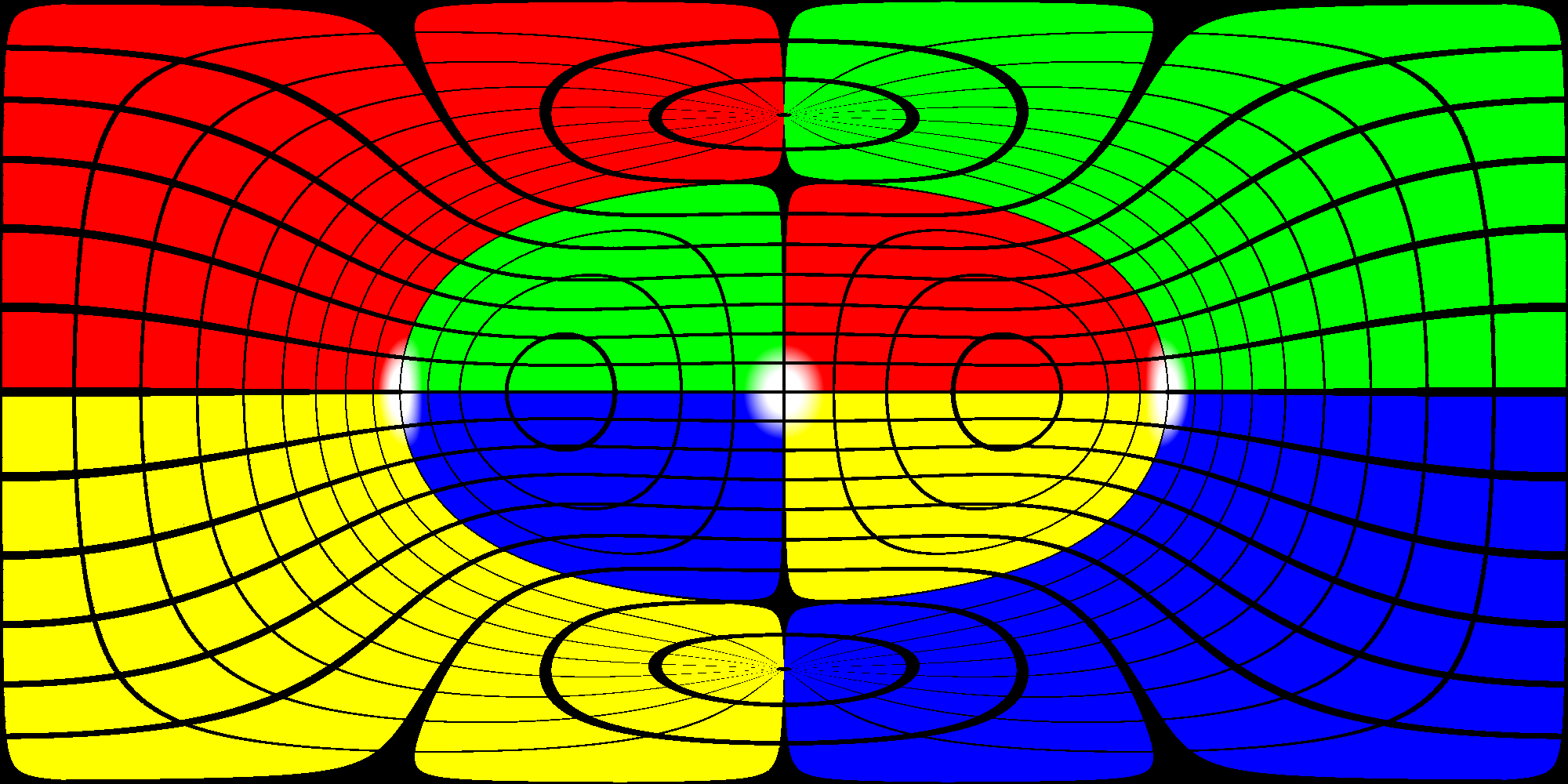}}
\caption{Panoramic images of the gravitational lensing of the dilatonic Melvin geometry for different values of $a$. In this figure, the observer is located at the equatorial plane ($\theta=\pi/2$) and at the perimetral radius $Br_p=1$. The radius of the celestial sphere is chosen as $Br_{\rm cs}=2.5$.} %
\label{lensing_melvind}
\end{figure}

Before exploring the shadow and gravitational lensing in SdM, let us first analyze the gravitational lensing in the dilatonic Melvin geometry ($M=0$). In Fig.~\ref{lensing_melvind} we show the panoramic images for the gravitational lensing of the dilatonic Melvin geometry with different values of $a$. In this figure, the observer is located at $B\,r_p=1$, $\theta_{\rm obs}=\pi/2$ and the celestial sphere has a radius $B\,r_{\rm cs}=2.0$. We have chosen two values of $a$ for which a stable LR exists and two values for which it does not exist. One key difference between the panoramic images for $a<\sqrt{3}$ and $a\geqslant \sqrt{3}$ is the existence of chaotic lensing regions for the former case. Such chaotic lensing regions for $a<\sqrt{3}$ arise due to the existence of bound orbits, $i.e.$ orbits that are bound between two radii and hence cannot escape to infinity. This is closely connected to the existence of a stable LR. For $a \geqslant \sqrt{3}$ the gravitational lensing becomes regular, since light rays can escape to infinity and there are no bound orbits in general, as the stable LR is absent.  

\subsection{Shadow and gravitational lensing for the SdM BH: Case I}
\begin{figure}
\includegraphics[scale=0.68]{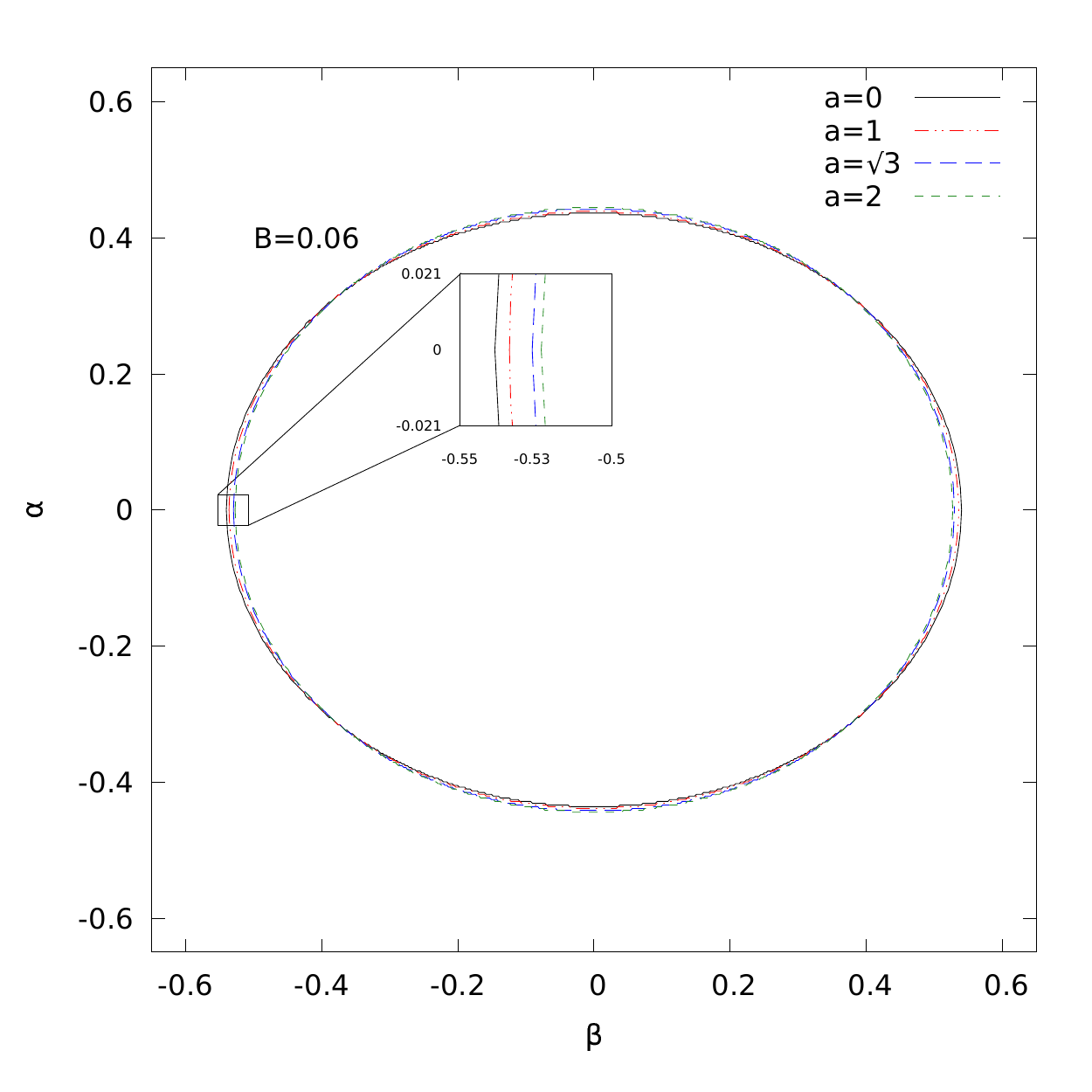}
\caption{Contour of the shadow of SdM BHs for several values of $a$ and fixed $BM=0.06$. In this figure, the observer is located at the equatorial plane ($\theta=\pi/2$) and at the perimetral radius $r_p=10M$.}
\label{shadow_edge}
\end{figure}

\begin{figure*}
\centering
\subfigure[$a=0$]{\includegraphics[scale=0.2]{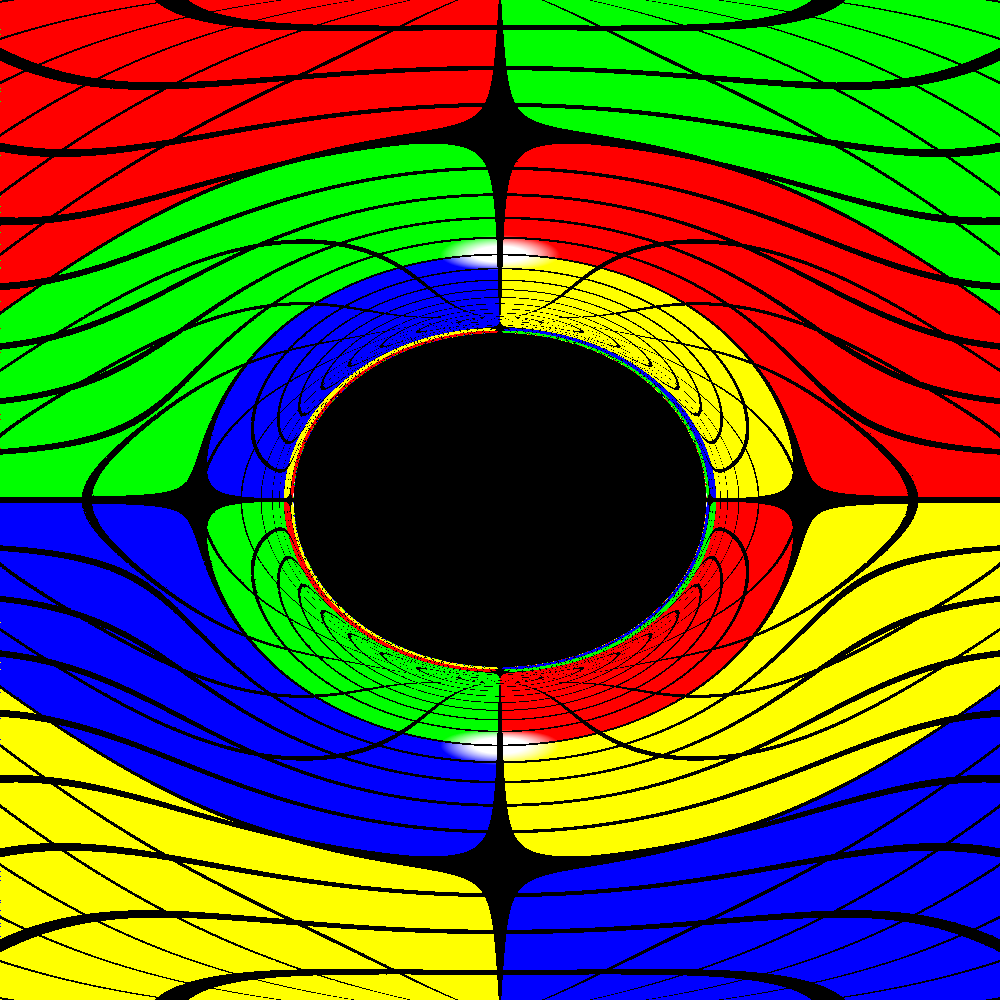}}
\subfigure[$a=1$]{\includegraphics[scale=0.2]{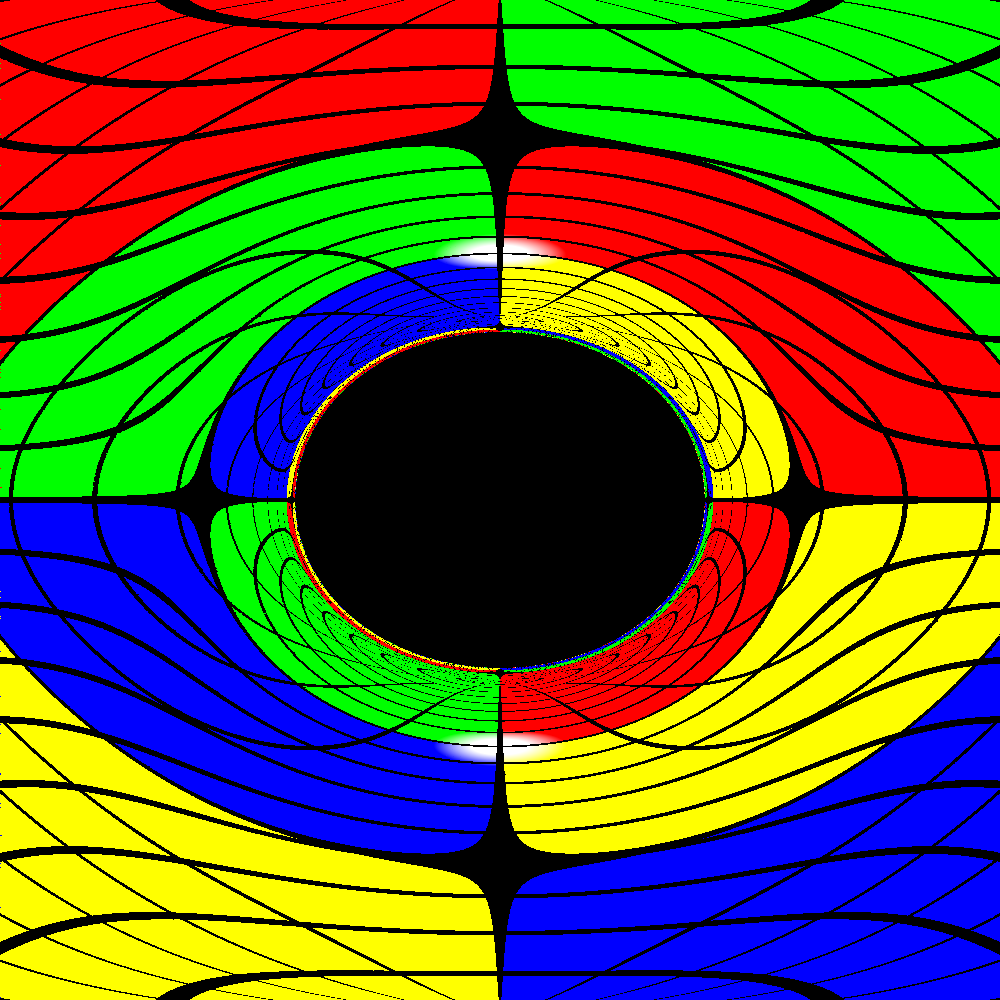}}
\subfigure[$a=\sqrt{3}$]{\includegraphics[scale=0.2]{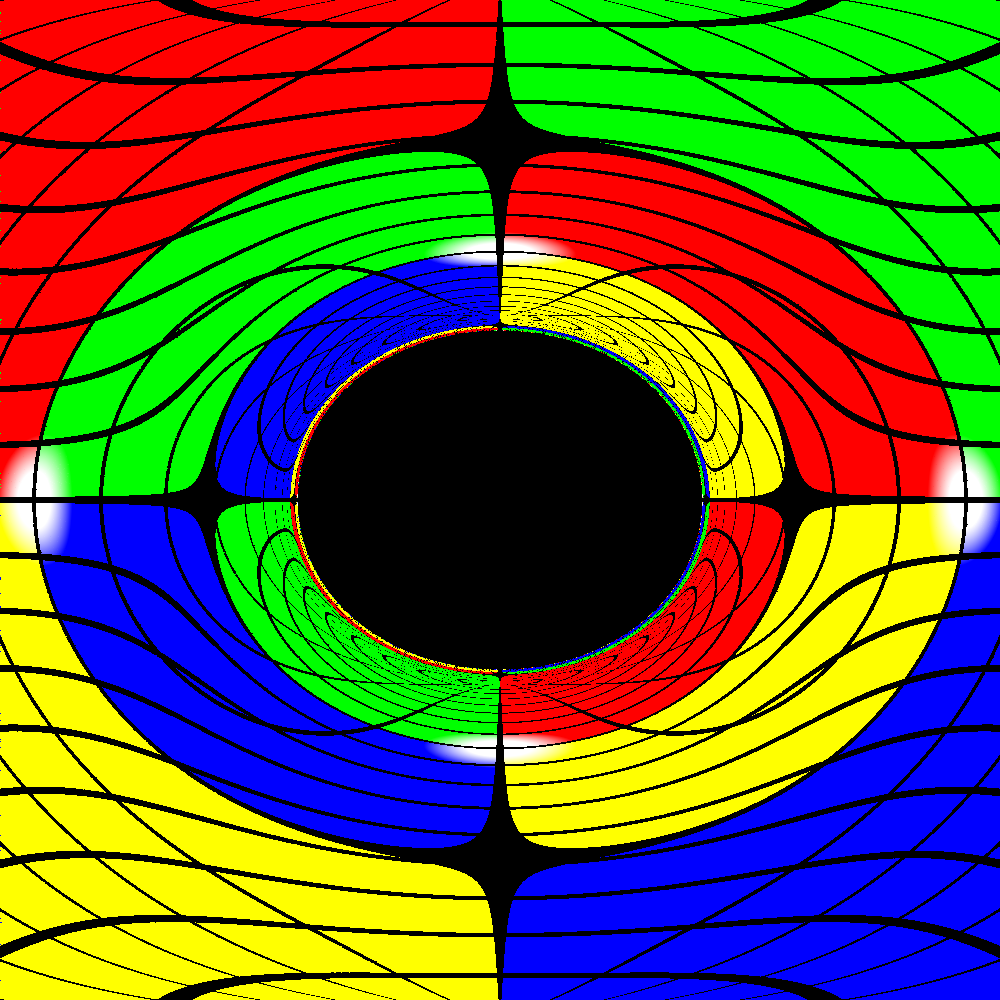}}
\subfigure[$a=2$]{\includegraphics[scale=0.2]{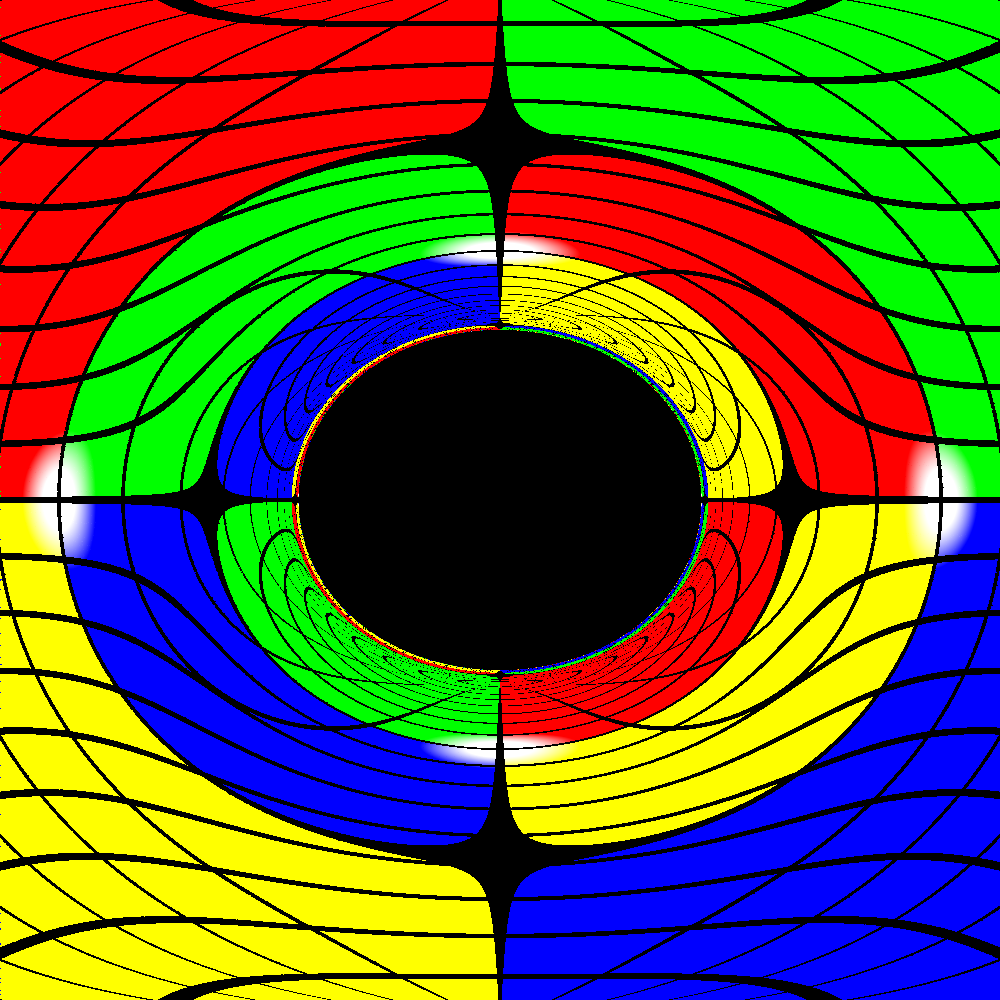}}
\caption{Shadow and gravitational lensing of SdM BHs for $BM=0.06$ and different values $a$. In this figure, the observer is located at the equatorial plane ($\theta=\pi/2$) and at the perimetral radius $r_p=10M$.} %
\label{shadow_partial_view}
\end{figure*}

Let us now analyze the shadows and gravitational lensing of the SdM BH. Firstly, we compare the shadow and gravitational lensing for the three distinct cases (Case I, II-A and II-B) with $BM \ll 1$, which, at least in the vicinity of the horizon, may be closer to real astrophysical environments. In Fig.~\ref{shadow_edge} we show solely the contour of the shadow for several values of $a$, and fixed $BM=0.06$. We place the observer at the equatorial plane ($\theta=\pi/2$) and at the perimetral radius $r_p=10M$. This shows that the shadow edge varies weakly with $a$. 
In Fig.~\ref{shadow_partial_view} we show the shadow and also the gravitational lensing for $BM=0.06$ and the same values of $a$ as in Fig.~\ref{shadow_edge}. From Figs.~\ref{shadow_edge} and \ref{shadow_partial_view}, we note that the contour of the shadow has an oblate shape in comparison to the Schwarzschild case, and as we increase the value of $a$, keeping fixed the value of $BM$, the oblateness slightly decreases. The same applies to some lensing features in the images: they become less oblate when $a$ increases.

\begin{figure}
\includegraphics[scale=0.12]{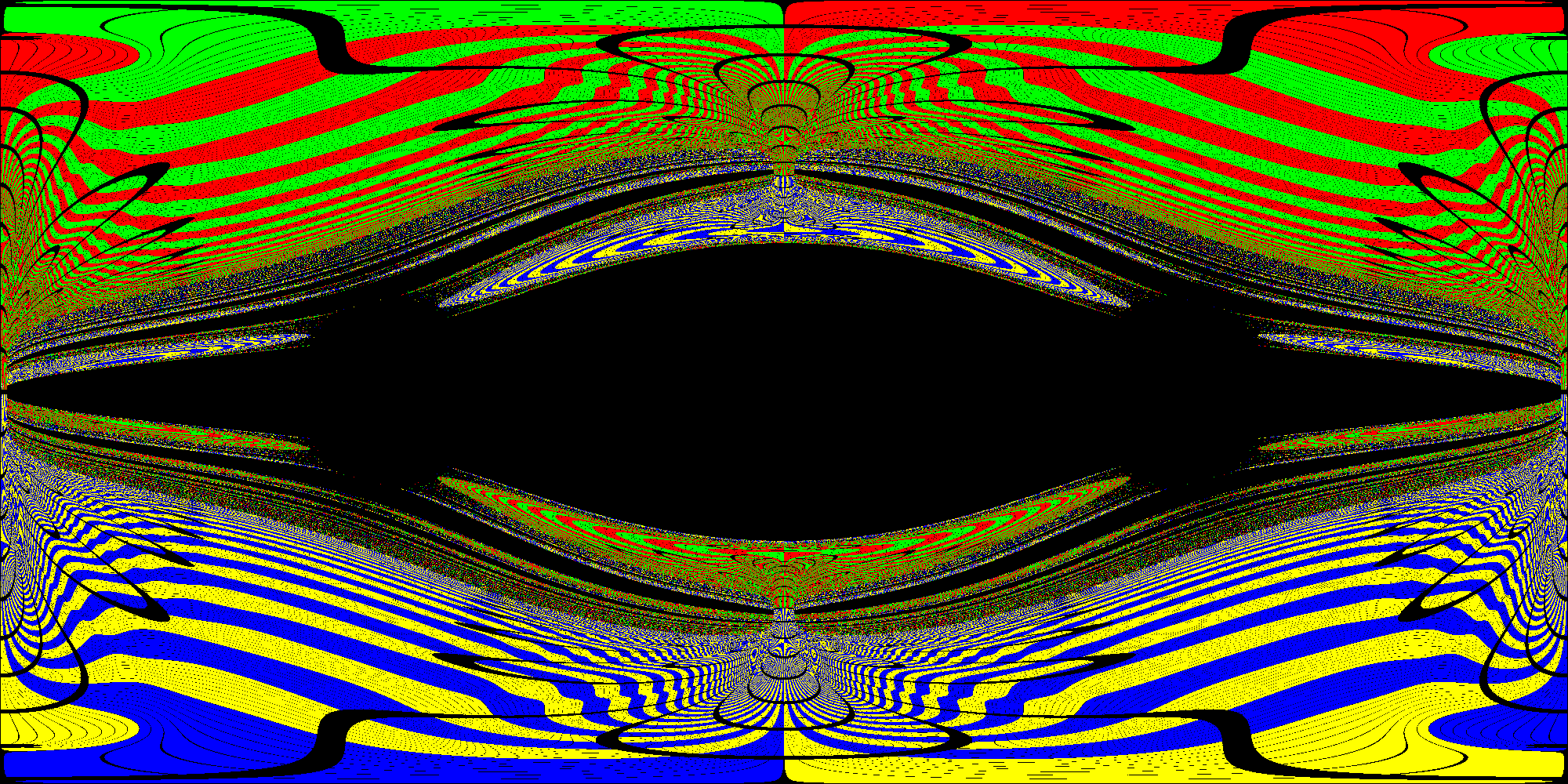}
\caption{Panoramic image of the shadow and gravitational lensing for the SdM BH with $BM=0.2$ and $a=0.5$ (case I). The observer is located at the perimetral radius $r_p=5M$ and $\theta_{\rm obs}=\pi/2$.}
\label{panoramic_shadow_a1}
\end{figure}

In Fig.~\ref{panoramic_shadow_a1} we show the panoramic image of the shadow and gravitational lensing for a SdM BH with $BM=0.2$ and $a=0.5$ (case I). We have chosen the observer to be located at the perimetral radius $r_p=5M$ and at the equatorial plane ($\theta_{\rm obs}=\pi/2$). We observe that the shadow and gravitational lensing display a chaotic behavior under this stronger external magnetic field. Since there are no LRs along the equatorial plane, the shadow is panoramic for $BM=0.2$ and $a=0.5$. The chaotic behavior for the shadow and gravitational lensing in Fig.~\ref{panoramic_shadow_a1} arises due to the confining behavior of the geometry in Case I. Since light rays cannot escape to infinity, they can bounce at several turning points before falling into the event horizon or reaching the celestial sphere. This sort of ``ping-pong" leaves a chaotic pattern imprint, alongside the panoramic shadow in the SdM solution in this parameter range, as in the $a=0$ case~\cite{MSBH21}.

\subsection{Shadow and gravitational lensing for the SdM BH: Case IIA}
\begin{figure}
\includegraphics[scale=0.6]{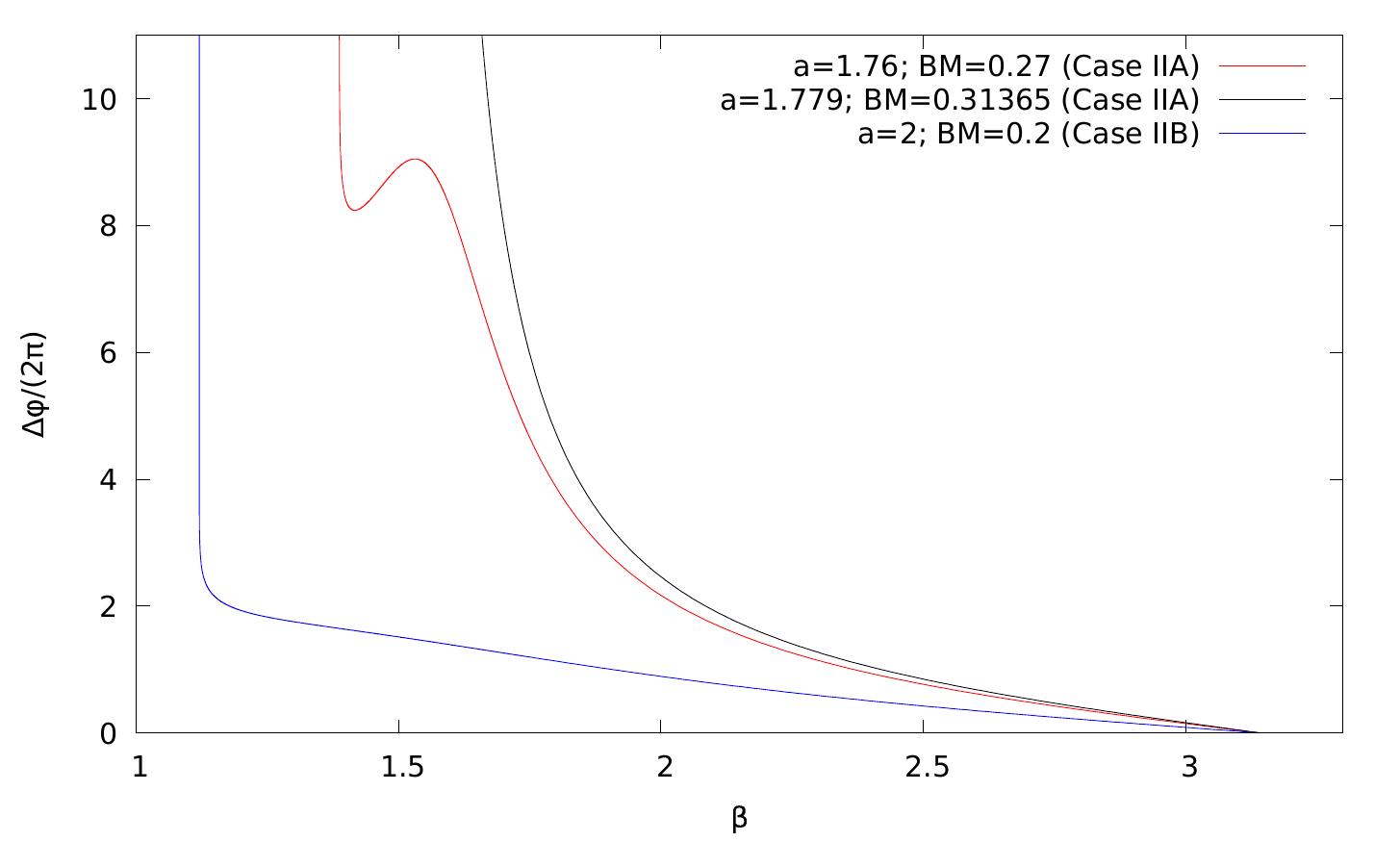}
\caption{Scattered angle for null geodesics constrained to the equatorial plane for different values of  $a$ and $BM$. In this figure, we have selected two SdM BHs corresponding to case IIA and one SdM BH corresponding to case IIB. We notice that when two unstable LRs are present, there may be a local maximum in the scattered angle.}
\label{scattered_angle}
\end{figure}

\begin{figure}
\centering
\subfigure[$a=1.76$]{\includegraphics[scale=0.12]{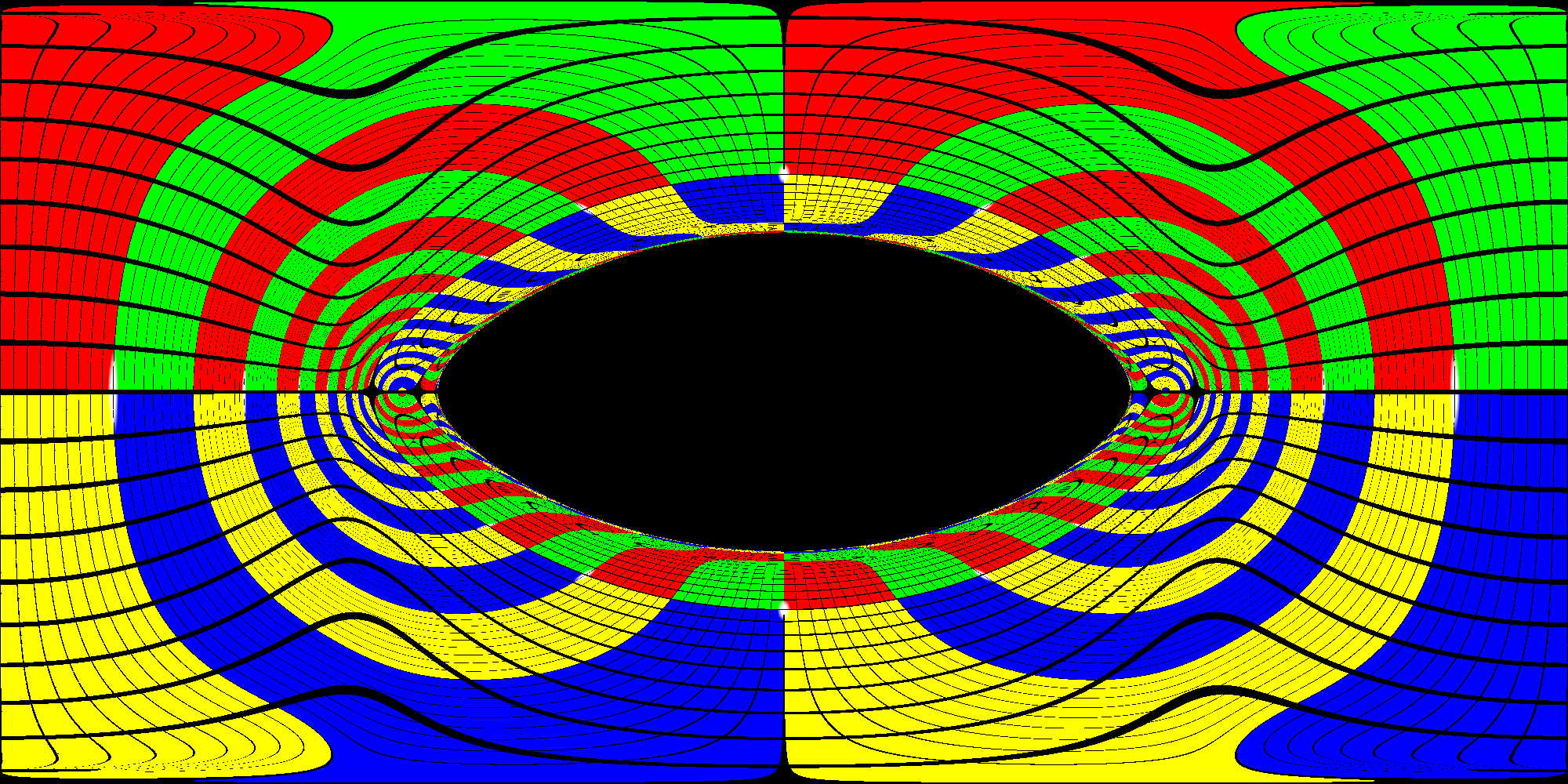}}
\subfigure[$a=1.779$]{\includegraphics[scale=0.12]{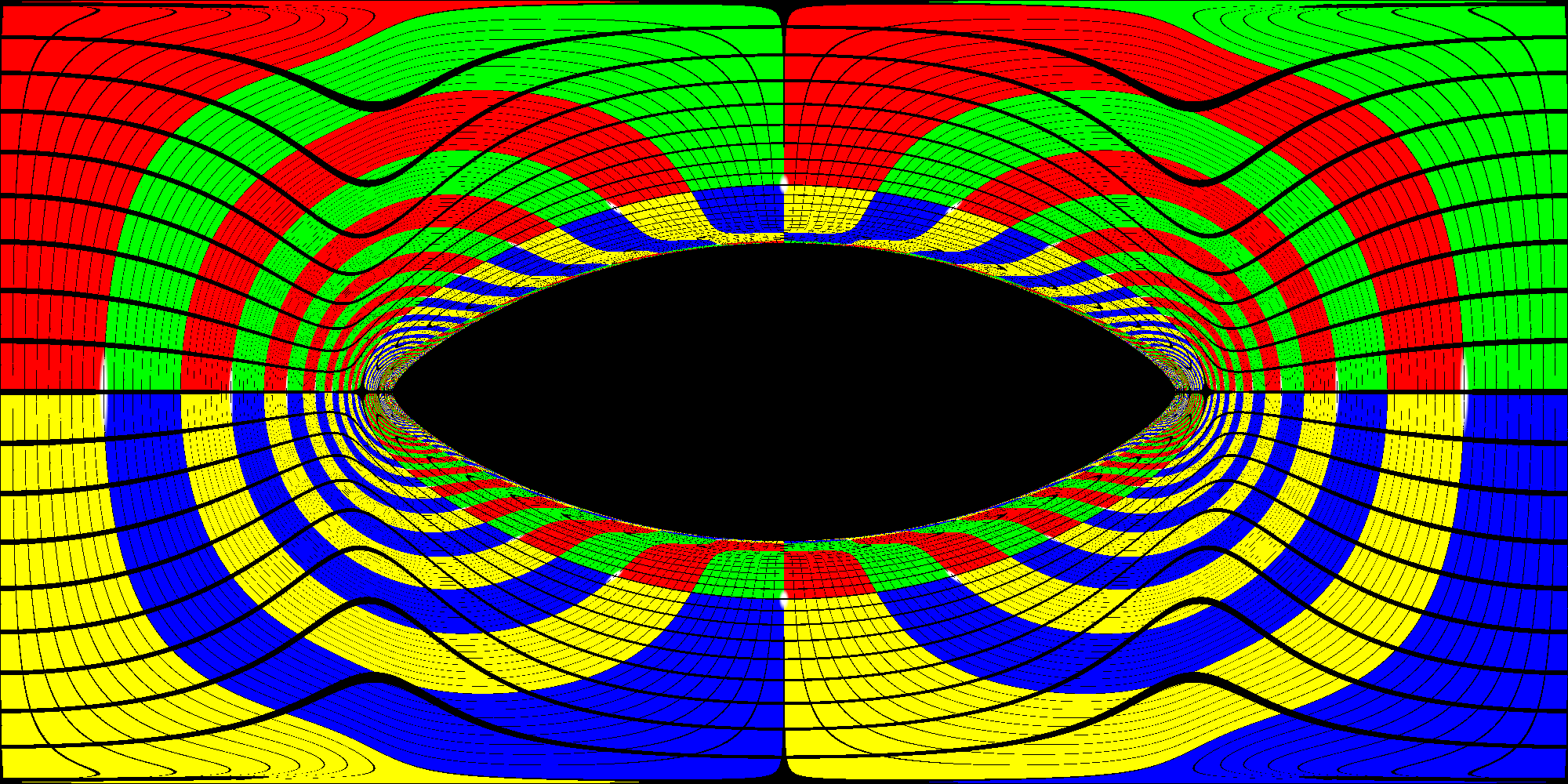}}
\caption{Panoramic image of the shadow and gravitational lensing for two SdM BHs within case II. Top panel: $BM=0.27$ and $a=1.76$. Bottom panel:  $BM=0.31365$ and $a=1.779$. The observer is located at the equatorial plane ($\theta_{\rm obs}=\pi/2$) and at the perimetral radius $r_p=5M$.} %
\label{panoramic_shadow_caseIIA}
\end{figure}

In Fig.~\ref{panoramic_shadow_caseIIA} we show the panoramic image of the shadow and gravitational lensing for a SdM BH with $BM=0.27$ and $a=1.76$ (top panel) and $BM=0.31365$ and $a=1.779$ (bottom panel), corresponding to Case IIA. The perimetral radius and the $\theta$ coordinate of the observer were chosen to be the same as in Fig.~\ref{panoramic_shadow_a1}. From Fig.~\ref{panoramic_shadow_caseIIA} we note that the shadow and gravitational lensing for case IIA is quite distinctive from case I for large values of $MB$. In particular, we note that there are no chaotic regions in the panoramic images and the shadows are not panoramic. The absence of chaotic regions and panoramic shadows for case IIA arises from the fact that light rays can escape to infinity (see Eq.~\eqref{H_infinity}) and do not describe bound orbits with several turning points. From Fig.~\ref{panoramic_shadow_caseIIA} we note the existence of strong gravitational lensing far from the shadow edge in the panoramic images. This  is caused mainly by the strong external magnetic field, which plays a role even for large values of the radial coordinate $r$. We have chosen $BM=0.27$ and $a=1.76$ (top panel) because there are two unstable LRs (and one stable) on the equatorial plane for such values, and the observer is located between the unstable LRs.  On the other hand, for $BM=0.31365$ and $a=1.779$ (bottom panel) the observer is located outside the LRs. We note a slight difference on the scattering next to the shadow edge on the equatorial plane. The existence of a second unstable LR may give rise to a local maximum on the scattered angle for light rays constrained to the equatorial plane, as can be seen in Fig.~\ref{scattered_angle} where we compare the scattering angle for light rays constrained to the equatorial plane for cases IIA and IIB. The LR that determines the shadow edge at the equatorial plane is the one with the smallest value of $\eta_{LR}$, the so-called dominant LR~\cite{DegeneratedShadow:2021}. We point out that the shadow edge outside the equatorial plane is determined by fundamental photon orbits (FPO), and the existence of two unstable LRs at the equatorial plane does not necessarily imply the existence of more than one FPO for two given impact parameters ($\alpha,\beta$).

\begin{figure}
\centering
\subfigure[$a=2$]{\includegraphics[scale=0.12]{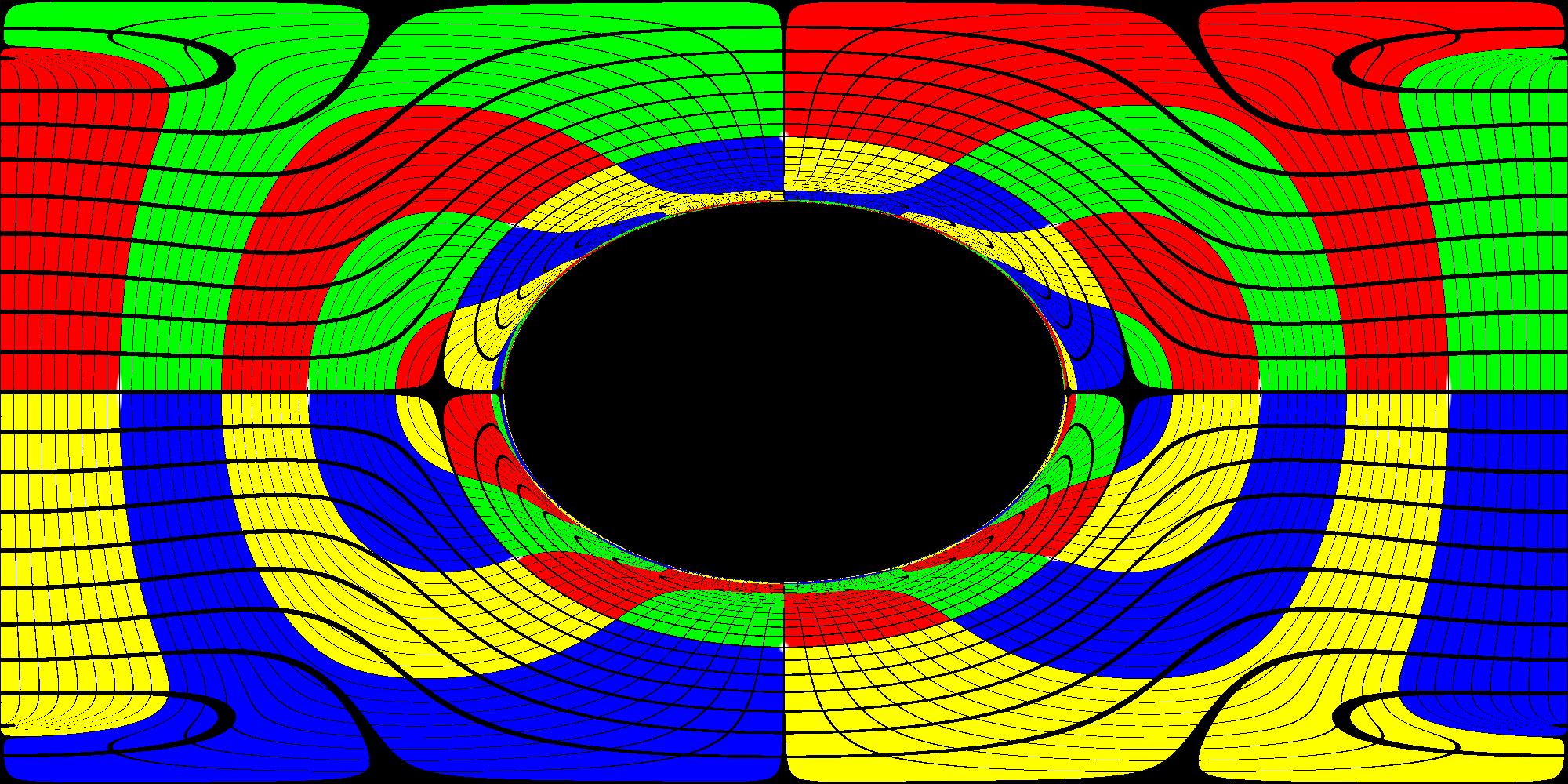}}
\caption{Panoramic image of the shadow and gravitational lensing for the SdM BH with $BM=0.2$ and $a=2$ (case II-B). In this figure, the observer is located at the equatorial plane ($\theta_{\rm obs}=\pi/2$) and at the perimetral radius $r_p=5M$.} %
\label{panoramic_shadow_caseIIB}
\end{figure}

\subsection{Shadow and gravitational lensing for the SdM BH: Case IIB}
In Fig.~\ref{panoramic_shadow_caseIIB} we  show the panoramic image of the shadow and gravitational lensing for a SdM BH with $BM=0.2$ and $a=2$, which is an example of case IIB. The perimetral radius and the polar angle of the observer are the same as in Fig.~\ref{panoramic_shadow_a1} and \ref{panoramic_shadow_caseIIA}. We note that, as in case IIA, the shadow and gravitational lensing are regular, and do not display chaotic regions. We also note the existence of gravitational lensing far from the shadow edge due to the influence of the external magnetic field for large values of the radial coordinate. Comparing case IIA and IIB, we note that as we increase the value of the parameter $a$, the oblateness of the shadow decreases becoming more round. We remark that cases IIA and IIB behave similarly to an asymptotically flat spacetime concerning the motion of null geodesics, despite the non-zero magnetic field extending all the way to spatial infinity.

\section{Final remarks}
\label{Final remarks}
We have considered the null geodesic flow in the SdM family of BH solutions of Einstein-Maxwell-dilaton theory,  where a non-minimal coupling between the Maxwell field and the dilaton exists, ruled by the parameter $a$. In particular, we investigated the LRs, total TC, shadows and gravitational lensing  for arbitrary values of the dilaton coupling. The present work generalizes the Einstein-Maxwell theory ($a=0$) results obtained in Ref.~\cite{MSBH21}, where the Schwarzschild solution immersed in an external magnetic field was considered. 

We analyzed how the dilaton coupling $a$ impacts the motion of null geodesics in the asymptotic region. We found that for $a\leqslant \sqrt{3}$, (non-radially moving) light rays cannot escape to infinity, which is a manifestation of the confining box behavior of the Melvin asymptotics~\cite{MSBH21}. For $a\geqslant \sqrt{3}$, however, the LR rays can reach spatial infinity and the SdM family loses the confining box behavior. Curiously, for the KK theory ($a=\sqrt{3}$), the light rays can escape to infinity only if the impact parameter is greater or equal to $1/|B|$. This analysis is valid both for the Melvin dilatonic ($M=0$) and the SdM BH solutions. These results show that the dilaton coupling modifies the asymptotic behavior of the spacetime, as we indeed confirmed by computing the LRs TC.

We computed the LRs in the Melvin dilatonic and SdM families of solutions and obtained that there is a qualitative change at $a=\sqrt{3}$ in both cases. For the Melvin dilatonic family with $a< \sqrt{3}$ there is a stable LR tube whose cylindrical coordinate depends on $a$ and $B$, and for $a \geqslant \sqrt{3}$ there is no LR in a finite radial coordinate. On the other hand, for the SdM family with $a\leqslant \sqrt{3}$ there may be zero or two LRs outside the horizon (case I), while for $a>\sqrt{3}$ there may be three (case IIA) or one LRs (case IIB). For cases I and IIA it is possible to have the creation/annihilation of a pair of LRs by fixing the dilatonic coupling and varying the magnetic field strength. Case IIB admits only one unstable LR, and as we increase the value of $B$ and fix $a$, the radial coordinate of the LR increases monotonically. By computing the TC of the SdM family we were able to shown that the dilaton coupling induces a \textit{topological transition}, $i.e.$ the TC changes discontinuously from TC=$0$ (case I) to TC=$-1$ (cases IIA and IIB). This \textit{topological transition} arises due to the modification in the asymptotic behavior of the SdM family. The horizon and the axis limit are not modified qualitatively by varying the dilaton coupling.

Taking advantage of the higher dimensional interpretation of the KK case $a=\sqrt{3}$ we were able to give an interpretation of the LRs of the SdM BH in $D=5$ in that case.

Finally, we have applied the backwards ray-tracing technique to the SdM family in order to study the gravitational lensing and shadows. For case I, the shadow and gravitational lensing present a chaotic behavior, while for the cases IIA and IIB they become regular even for high values of the magnetic field. Moreover, for case I it is possible to have panoramic shadows, as a consequence of the absence of LRs outside the horizon, similarly to the $a=0$ case investigated in Ref.~\cite{MSBH21}. Our results suggest that the shadows and gravitational lensing of the SdM BHs are qualitatively similar to those of an asymptotically flat spacetime for large values of the dilatonic coupling.

As future work, it could be interesting to analyse the possible impact of rotation in our study, as it was done in \cite{Wang:2021ara} for the case of $a=0$.

\acknowledgments
The authors thank Funda\c{c}\~ao Amaz\^onia de Amparo a Estudos e Pesquisas (FAPESPA),  Conselho Nacional de Desenvolvimento Cient\'ifico e Tecnol\'ogico (CNPq) and Coordena\c{c}\~ao de Aperfei\c{c}oamento de Pessoal de N\'{\i}vel Superior (Capes) - Finance Code 001, in Brazil, for partial financial support.
This work is supported by the Center for Research and Development in Mathematics and Applications (CIDMA) through the Portuguese Foundation for Science and Technology (FCT - Funda\c{c}\~ao para a Ci\^encia e a Tecnologia), references UIDB/04106/2020, UIDP/04106/2020. PC is supported by the Individual CEEC program 2020 funded by the FCT. JZY is supported by the China Scholarship Council. We acknowledge support  from the projects PTDC/FIS-OUT/28407/2017, CERN/FIS-PAR/0027/2019 and PTDC/FIS-AST/3041/2020. This work has further been supported by  the  European  Union's  Horizon  2020  research  and  innovation  (RISE) programme H2020-MSCA-RISE-2017 Grant No.~FunFiCO-777740.

\bibliography{MDBH}

\end{document}